\documentclass[12pt]{article}

\usepackage{latexsym,amsmath,amssymb,theorem,epsfig,multirow}
\usepackage[noconfig]{refstyle}
\usepackage[dvipsnames]{xcolor}
\usepackage[hyperfootnotes=false, linktocpage=true, colorlinks, citecolor=blue, linkcolor=blue, urlcolor=Maroon]{hyperref}
\usepackage[all]{xy}

\newlength{\xtrawidth}
\newlength{\xtraheight}
\DeclareFontFamily{OT1}{rsfs10}{}
\DeclareFontShape{OT1}{rsfs10}{m}{n}{ <-> rsfs10 }{}
\DeclareMathAlphabet{\mathscript}{OT1}{rsfs10}{m}{n}

\numberwithin{equation}{section}

\newcommand{\pt}{\partial}

\newcommand{\C}{\ensuremath{{\mathbb{C}}}}

\newcommand{\CP}{\ensuremath{\mathop{\null {\mathbb{P}}}}\nolimits}

\def\fnote#1#2{\begingroup\def\thefootnote{#1}\footnote{#2}
     \addtocounter{footnote}{-1}\endgroup}

\def\a{\alpha}
\def\b{\beta}

\def\d{\delta}

\def\l{\lambda}

\def\o{\omega}

\def\s{\sigma}
\def\t{\tau}

\def\gsim{ \lower .75ex \hbox{$\sim$} \llap{\raise .27ex \hbox{$>$}} }
\def\lsim{ \lower .75ex \hbox{$\sim$} \llap{\raise .27ex \hbox{$<$}} }
\def\be{\begin{equation}}
\def\ee{\end{equation}}
\def\bea{\begin{eqnarray}}
\def\eea{\end{eqnarray}}

\def \td {\tilde}

\def \a {\alpha}

\def \a {\alpha }
\def \b {\beta }
\def \o{\omega }

\def \l {\lambda}

\def \ed {\end{document}}

\begin{document}

\begin{titlepage}

\title{{\LARGE\bf   Non-vanishing  Superpotentials in Heterotic String Theory  and Discrete Torsion}\\[1em] }
\author{
Evgeny I. Buchbinder${}^{1}$,
Burt A. Ovrut ${}^{2}$
}

\date{}
\maketitle
\begin{center} { 
${}^1${\it School of Physics (M013) \\ The University of Western Australia\\
35 Stirling Highway, Crawley WA 6009, Australia\\[3mm]
${}^2$Department of Physics and Astronomy \\
University of Pennsylvania \\
209 South 33rd Street, Philadelphia, PA 19104-6395, USA 
}}\\
\end{center}

\fnote{}{evgeny.buchbinder@uwa.edu.au,~ ~ovrut@elcapitan.hep.upenn.edu}

\vskip 0.5cm

\begin{abstract}
\noindent

We study the non-perturbative superpotential in $E_8 \times E_8$ heterotic string theory on a non-simply connected Calabi-Yau 
manifold $X$, as well as on its simply connected covering space $\tilde{X}$.
The superpotential is induced by the string wrapping  holomorphic, isolated, genus 0 curves. 
According to the residue theorem of Beasley and Witten, the non-perturbative  superpotential must vanish in a large class 
of heterotic vacua because the contributions from  curves in the same homology class cancel each other. 
We point out, however, that in certain cases the curves treated in the residue theorem as lying in the same homology class, 
can actually have different area with respect to the physical Kahler form and can
be in different homology classes.
In these cases, the residue theorem is not directly applicable and the structure of the superpotential is more subtle. 
We show, in a specific example, that the superpotential is non-zero both on $\tilde{X}$ and on $X$.
On the non-simply connected manifold $X$, we explicitly compute the leading contribution to the superpotential
from all holomorphic, isolated, genus 0 curves with  minimal area. The reason for the non-vanishing of the superpotental on $X$ is that the second homology class 
contains a finite part called discrete torsion. As a result, the curves with the same area are distributed among different torsion classes
and, hence, do not cancel each other. 

\end{abstract}

\thispagestyle{empty}

\end{titlepage}

\tableofcontents

\section{Introduction}


Compactification of $E_{8} \times E_{8}$ heterotic string theory on smooth Calabi-Yau (CY) threefolds can lead to realistic particle physics models. For example, heterotic M-theory vacua consisting of stable, holomorphic $SU(4)$ vector bundles defined by ``extension'' over a class of Schoen CY threefolds can produce exactly the spectrum of the minimal supersymmetric standard model (MSSM) with gauged $B-L$ symmetry~\cite{Braun:2005ux, Braun:2005bw, Braun:2005zv, Braun:2005nv}. Similarly, heterotic M-theory compactified on other classes of CY threefolds, such as the tetra-quadric, carrying ``monad'' vector bundles can lead to the MSSM at low energy, with or without gauged $B-L$ symmetry~\cite{Anderson:2007nc, Anderson:2009mh, Buchbinder:2013dna, Buchbinder:2014qda, Buchbinder:2014sya}. Although these string vacua realize the correct spectrum and interactions of low energy particle physics, there remains a fundamental problem; that is, that the associated threefolds and vector bundles have moduli that generically have no potential energy. Therefore, the vacuum values of these fields can be dynamically unstable and, even if time-independent, cannot be uniquely specified--thus rendering explicit predictions of the values of supersymmetry breaking and physical parameters impossible.  It follows that the stabilization of both geometric and vector bundle moduli is one of the most important problem in heterotic string theory. 

A non-vanishing potential energy for the geometric moduli, that is, the complex structure~\cite{Anderson:2010mh} and Kahler moduli, can occur for specific heterotic string vacua due to both perturbative and non-perturbative effects. This leads to partial, and in some toy cases complete, stabilization of these moduli~ \cite{Anderson:2011cza}. However, the situation for vector bundle moduli is more difficult. Here, there is no perturbative contribution to their potential energy and one must examine possible non-perturbative effects.
A non-perturbative superpotential can, in principal, be generated by string instantons~\cite{Dine:1986zy, Dine:1987bq, Becker:1995kb, Witten:1996bn, Donagi:1996yf, Witten:1999eg, Harvey:1999as, Lima:2001jc, Lima:2001nh}. 
It  depends (inversely) exponentially  on the Kahler moduli, 
and also contributes to a potential energy for both complex structure and, importantly, the vector bundle moduli through 1-loop determinants. 
However, it is difficult to compute these 1-loop quantities. So far, this has only been carried out for  
specific examples of
elliptically fibered CY threefolds with spectral cover vector bundles~\cite{Buchbinder:2002ic, Buchbinder:2002pr}. It is important, therefore, 
to generalize these constructions to  more realistic vacua, such as those mentioned above. Even then, to find the complete superpotential one has to sum up 
the contributions from all holomorphic, isolated, genus 0 curves. 
Beasley and Witten showed that, in a large class of models, these contributions cancel against each other~\cite{Beasley:2003fx, Beasley:2005iu}.
Hence, in addition to calculating the instanton generated superpotential for specific curves in more realistic vacua, one must then show that these contributions do not cancel each other; that is, that the Beasley-Witten theorem is not applicable to these theories. In this paper, we take a first step in that direction by explicitly calculating the complete leading order instanton superpotential for  a heterotic vacuum consisting of a Schoen~\cite{{Schoen}} threefold geometry and a simple ``extension'' $SU(3)$ vector bundle--similar, but not identical, to the heterotic standard model in~\cite{Braun:2005nv}. Although our Schoen threefold is a complete intersection CY manifold (CICY) and the vector bundle descends from a vector bundle on the ambient space-- two of the three main conditions required by the Beasley-Witten theorem--we find in this theory that the Beasley-Witten theorem is not applicable and that the superpotential indeed does not vanish. Extending this work to exact heterotic standard model vacua will be carried out elsewhere.

We start our analysis with a theory on a Schoen threefold $\td{X}$
which is a CICY in the ambient space ${\cal A}= {\mathbb P}^1 \times {\mathbb P}^2 \times {\mathbb P}^2$. We will also consider only those vector bundles $\td{V}$ on $\td{X}$ that descend from a vector bundle $\cal{V}$ on $\cal{A}$.
These vacua satisfy two of the three conditions of the residue theorem of Beasley and Witten and, therefore, one might expect the complete non-perturbative superpotential  to vanish. However, we point out that the Beasley-Witten residue theorem
additionally assumes that the area of all holomorphic curves on the CICY is computed using the restriction of the Kahler form on the ambient space. 
Usually, this restriction does give the complete Kahler form on the CY manifold--but there are cases when it does not. 
These more subtle cases arise when the CICY manifold has more $(1, 1)$ classes than does the ambient space. 
As a result, curves which have the same area with respect to the restriction of the Kahler form of the ambient space,
can actually have different area with respect to the true  Kahler form on the Calabi-Yau space and, hence, can lie in different homology classes. 
The Schoen manifold studied in the paper has this property. It has 19 $(1, 1)$ classes whereas the ambient space has only 3. 
We show that there are holomorphic, isolated, genus 0 curves in this manifold which are unique in their  homology classes
despite having the same area with respect to the restriction of the Kahler form of ${\cal A}$. 
Thus, for an arbitrary vector bundle the contributions to the non-perturbative superpotential due to these curves cannot cancel 
each other because they are weighted with different area. This way, one can get around the Beasley-Witten residue theorem. 

Furthermore, our CICY Schoen threefold is chosen to have a freely acting ${\mathbb Z}_3 \times {\mathbb Z}_3$ symmetry group. 
We then mod our this discrete action, to obtain  a non-simply connected Calabi-Yau space with $\pi_1 = {\mathbb Z}_3 \times {\mathbb Z}_3$. 
For a toy choice of  a vector bundle which descends from the ambient space, the non-perturbative superpotential for all holomorphic, isolated, genus 0 curves
with minimal area is computed. For simplicity, we perform our calculations for a fixed complex structure. Hence, the only 1-loop determinant
which needs to be computed is the Pfaffian of the Dirac operator on these curves. Since we do not know either the metric or the gauge connection, 
we use an algebraic method (similar to the one developed in \cite{Buchbinder:2002ic, Buchbinder:2002pr}) to compute the Pfaffians. 
They turn out to be homogeneous, degree 2 polynomials on the moduli space of 
vector bundles. We show that the sum of the contributions from these curves is non-zero. Here, the main reason 
for the non-vanishing of the superpotential is the discrete part of the second homology group, called discrete torsion. 
Due to torsion, curves which have the same area actually lie in different classes of the second homology group with integer coefficients. 
These different classes are labeled by the characters of the torsion subgroup--which in the present case is ${\mathbb Z}_3 \oplus {\mathbb Z}_3$.
Hence, in this case the non-vanishing of the superpotential can also be attributed to existence of holomorphic, isolated, genus 0 curves which 
are unique in their integral homology classes. 

The paper is organized as follows. In Section 2, we start with reviewing the structure of the non-perturbative superpotential
in heterotic string theory, mostly following~\cite{Witten:1999eg}. Then we review the residue theorem of Beasley and Witten, 
pointing out that it is directly applicable only when the Kahler form on the Calabi-Yau manifold is the restriction of the Kahler form of the ambient space. 
We also discuss how the structure of the superpotential is modified if the second homology group with integer coefficients
contains discrete torsion. Holomorphic, isolated, genus 0 curves which are in the same  real homology classes and which, hence, have 
the same area, are distributed among different torsion classes labeled by the characters of the torsion group. 
These characters arise as extra factors in the superpotential. In Section 3, we review some mathematical propeties of the Schoen manifold
and of its quotient by ${\mathbb Z}_3 \times {\mathbb Z}_3$. We also review the type II prepotential on the quotient computed 
in~\cite{Braun:2007tp, Braun:2007xh, Braun:2007vy}. We point out that there are 9 holomorphic, isolated, genus 0 curves 
of the quotient having the same minimal area, but each lying in its own torsion class. 
In Section 4, we show that the pre-image of these curves 
on the covering Schoen manifold consists of 81 curves lying in 81 different homology classes. 
These curves have different area due to the Kahler classes which are non-invariant under ${\mathbb Z}_3 \oplus {\mathbb Z}_3$
and, hence, do not descent to the quotient. 
In particular, this implies that the superpotential on the Schoen manifold is non-zero. In the remaining part of the paper, we
compute the superpotential on the quotient manifold due to the above 9 isolated curves with minimal area. In this paper, we do the calculation 
for a toy model. More realistic vacua will be discussed elsewhere. 
In Section 5, we construct a toy vector bundle with structure group $SU(3)$ which descends to the quotient manifold. 
We show that its moduli space is a projective space and find its explicit parametrization. 
In Section 6, we compute the Pfaffian of the Dirac operator on the curves of interest. On each curve, the result is a
homogeneous polynomial which we find explicitly up to an overall coefficient. We show that, since all curves are in different integral 
homology classes, their contributions pick up different torsion factors and, hence, they do not cancel each other. 
In the Conclusion, we summarize our results and discuss directions for further research. 
Finally, Appendices A, B, C and D are devoted to discussing various technical details.


\section{Non-perturbative superpotentials in heterotic string theory}
\label{super}


\subsection{The general structure of non-perturbative superpotentials}
\label{general}

We consider $E_8 \times E_8$ heterotic string theory compactified to four-dimensions on a  Calabi-Yau threefold X. As was extensively studied in a variety of 
contexts and papers~\cite{Dine:1986zy, Dine:1987bq, Becker:1995kb, Witten:1996bn, Donagi:1996yf, Witten:1999eg, Harvey:1999as, Lima:2001jc, Lima:2001nh, Buchbinder:2002ic, Buchbinder:2002pr},
the effective low-energy field theory may, in principle, develop a non-perturbative superpotential for the moduli fields 
generated by worldsheet/worldvolume instantons. The structure of the instantons
is slightly different in the weakly and strongly coupled heterotic string theories. Be that as it may, the superpotential has the same generic form. For concreteness,
we will discuss the weakly coupled case where the superpotential is generated by strings wrapping holomorphic, isolated, genus $0$
curves in X.\footnote{Holomorphic, non-isolated and/or higher genus curves contribute to higher order F-term interactions~\cite{Beasley:2005iu}.} Furthermore, for simplicity, we will restrict our discussion to the ``observable'' sector; that is, to the superfields associated with the first $E_{8}$ factor of the gauge group.The superpotential is then determined by the
classical Euclidean worldsheet action $S_{cl}$
evaluated on the instanton solution and by the 1-loop determinants of the fluctuations around this solution. 
Let $C$ be a holomorphic, isolated, genus $0$ curve in X. 
Then the general form of the superpotential induced by a string wrapping $C$ 
is~\cite{ Witten:1999eg}
\be 
W (C)= {\rm exp}\Big[ -\frac{A(C)}{2 \pi \a'} + i \int_C B \Big]
\frac{{\rm Pfaff}  ({\bar \pt}_{{\rm V}_C (-1)})  }{ [ {\rm det}'  ({\bar \pt}_{{\cal O}} )]^2  {\rm det} ({\bar \pt}_{NC} )}\,. 
\label{1.1}
\ee
Let us review various ingredients in this formula. The expression in the exponent is the classical Euclidean action evaluated on $C$. 
In the first term, $A(C)$, is the area
of the curve given by 
\be 
A(C)= \int_C \omega\,, 
\label{1.2}
\ee
where $\omega$ is the Kahler form on X. In the second term, $B$ is the heterotic string $B$-field which, in this expression, can be taken 
to be a closed 2-form, $d B=0$. Let $\omega_I$ be a basis of $(1,1)$-forms on X, $I=1, \dots, h^{1,1}$. Then we can expand
\be 
\o=\sum_{I=1}^{h^{1,1}} t^I \o_I\,, \qquad B=\sum_{I=1}^{h^{1,1}} \phi^I \o_I\,. 
\label{1.3}
\ee
Let us define the complexified Kahler moduli 
\be 
T^I= \phi^I +i \frac{t^I}{2 \pi \a'}\,. 
\label{1.4}
\ee
Then the exponential prefactor becomes
\be 
e^{i \a_I(C) T^I}\,,  \qquad  \a_I(C) =\int_C \o_I\,. 
\label{1.5}
\ee
By construction ${\rm Re} (i \a_I(C) T^I) <0$. 

Note that the exponential factor in~\eqref{1.1} can also be understood as a map from the curve $C$ to the non-zero complex numbers ${\mathbb{C}}^{*}$. That is,
\be
C \to  {\rm exp}\Big[ -\frac{A(C)}{2 \pi \a} + i \int_C B \Big]\,. 
 \label{1.55}
 \ee
Since the value of the integrals depends only on the homology class of the curve, the map is more appropriately expressed as
\be 
e^{-S_{cl}}: H_2 ({\rm X}, {\mathbb Z}) \to {\mathbb C}^*\, .
\label{1.6}
\ee
However, here there is an important caveat. 
In eqs.~\eqref{1.3}, \eqref{1.6}  we are assuming that the moduli space of the $B$-field is connected. As we will discuss below, this 
is not necessarily the case. Hence, the map~\eqref{1.6} needs to be refined.

Let us now discuss the 1-loop determinants. The first determinant is the Pfaffian of the Dirac operator 
which comes from integrating over the right moving fermions in the worldsheet theory. In heterotic compactifications, we have to specify 
the internal gauge field $A$ on X which satisfies the Hermitian Yang-Mills equations  
\be 
F_{m n}=0\,, \quad F_{\bar m \bar n}=0\,, \quad g^{m \bar n} F_{m \bar n}=0\, , 
\label{1.5.1}
\ee
where $m$ and ${\bar m}$ are holomorphic and anti-holomorphic indices on X and $g_{m {\bar n}}$ is the Ricci flat metric on X.
According to the theorem of Donaldson-Uhlenbeck-Yau, $A$ is a connection on a holomorphic polystable vector 
bundle V on X whose structure group is a subgroup of $E_8$. Then the Pfaffian in~\eqref{1.1} is the Pfaffian of the Dirac operator 
depending on the connection $A$ restricted to the curve $C$. Since the spin bundle on a genus $0$ curve is ${\cal O}_C (-1)$, 
we additionally  tensor V with ${\cal O}_C (-1)$ and denote $ {\rm V}_C (-1)= {\rm V}|_C \otimes {\cal O}_C (-1)$.
${\rm Pfaff}  ({\bar \pt}_{{\rm V}_C (-1)})$ depends on the moduli of the vector bundle V.
In principle, it can be explicitly expressed as a function of the gauge connection $A$ using the WZW model~\cite{Buchbinder:2002pr}. However, since no explicit solutions to the Hermitian Yang-Mills equations on X are known, 
it is unclear how to use this in practice. Since right moving worldsheet fermions are Weyl,  the ${\rm Pfaff}  ({\bar \pt}_{{\rm V}_C (-1)})$ is anomalous. However, this  
anomaly is cancelled by the variation of the $B$-field~\cite{Buchbinder:2002pr}. As the result, the Pfaffian of the Dirac operator 
is not a function on the moduli space of V but, rather, a section of some line bundle.
In the denominator in~\eqref{1.1}, ${\rm det} ({\bar \pt}_{NC})$ comes from integrating over bosonic fluctuations and is the determinant 
of the ${\bar \pt}$-operator on the normal bundle to the curve $C$. For an isolated, genus 0 curve, the normal bundle is  
$NC= {\cal O}_{C}(-1) \oplus {\cal O}_{C}(-1)$. Hence, ${\rm det} ({\bar \pt}_{NC}) =[ {\rm det} {\bar \pt}_{{\cal O}_C (-1)}]^2$. Finally, $ [ {\rm det}'  ({\bar \pt}_{{\cal O}} )]^2 $
is the ${\bar \pt}$-operator on the trivial line bundle which is a constant.

In general, a given homology class of X contains more than 1 holomorphic, isolated, genus $0$ curve. The number of 
these curves is referred to as to Gromov-Witten invariant. All such curves in the same homology class have the same area, the same classical action 
and the same exponential prefactor in~\eqref{1.1}. However, the 1-loop determinants, in general, are different. Hence, the contribution to the 
superpotential  from all curves $C_i$ in the homology class $[C]$ of the curve $C$ is given by (for simplicity, we remove the constant factor 
$[ {\rm det}'  ({\bar \pt}_{{\cal O}} )]^{-2}$)
\be 
W ([C])={\rm exp}\Big[ -\frac{A(C)}{2 \pi \a'} + i \int_C B \Big] \sum_{i=1}^{n_{[C]}}
\frac{{\rm Pfaff}  ({\bar \pt}_{{\rm V}_{C_i} (-1)})  }{[ {\rm det} {\bar \pt}_{{\cal O}_{C_i} (-1)}]^2}\,, 
\label{1.8}
\ee
where $n_{[C]}$ is the number of the holomorphic, isolated, genus $0$ curves in the homology class $[C]$.
To find the complete non-perturbative superpotential $W$, we then have to sum over all homology classes. That is,
\be
W=\sum_{[C] \in H_{2} ({\rm X})} W([C]) \ .
\label{new1}
\ee
%


\subsection{The residue  theorem of Beasley-Witten}
\label{BW}


In~\cite{Beasley:2003fx} (also see earlier papers~\cite{Distler:1986wm, Distler:1987ee, Silverstein:1995re, Basu:2003bq}) Beasley and Witten 
showed that, under some rather general assumptions, 
the sum~\eqref{1.8} must vanish for each homology class $[C]$. 
Here, we review their assumptions since they will be important later in the paper. Let $\tilde X$ be a complete intersection Calabi-Yau 
threefold in  the product of projective spaces\footnote{The results of Beasley and Witten are also expected to be valid for complete intersections 
in  toric varieties.} ${\cal A} = {\mathbb P}^{n_1} \times \dots \times  {\mathbb P}^{n_a}$. That is, $\tilde X$ is given by 
a set of  polynomial equations $p_1=0, \dots, p_m=0$ where $\sum_{i=1}^a n_i - m =3$. Additionally, assume that the Kahler form $\omega_{\td{X}}$ descends from the ambient space, that is, $\omega_{\td{X}} =\omega_{\cal{A}}|_{\td{X}}$, and that the vector bundle $\tilde V$ on $\tilde X$ 
is obtained as a restriction of a vector bundle ${\cal V}$ on ${\cal A}$, $\tilde V= {\cal V}|_{\tilde X}$. Then, it was shown by Beasley and Witten that if these assumptions are 
satisfied, the sum~\eqref{1.8} vanishes for any homology class. This result was proven in~\cite{Beasley:2003fx} and interpreted as a residue theorem. 

The proof in~\cite{Beasley:2003fx} is based on standard arguments of topological field theory and localization. First, they constructed a topological worldsheet 
action with target space ${\cal A}$ such that there exists a set of supersymmetric vacuum solutions--all with CICY threefold ${\tilde X}$, $\omega_{\td{X}}=\omega_{\cal{A}}|_{\td{X}}$ and ${\tilde V}= {\cal V}|_{\tilde X}$. Each such vacuum is associated with a holomorphic curve $C \subset {\tilde X}$. By the standard 
arguments of topological field theory, the correlation functions in this theory do not depend on the coupling. In one limit of the coupling, the correlators
are localized on this set of supersymmetric vacua solutions. This leads to eq.~\eqref{new1} for the total superpotential, where one uses the fact that non-isolated and/or higher genus curves only contribute to higher F-term interactions,. In another limit, the same correlators vanish because of unsaturated fermionic zero modes. Hence, $W=0$. Since the exponential factor is different for each homology class, Beasley and Witten concluded that the sum~\eqref{1.8} vanishes for  any homology class. 
The assumptions of Beasley and Witten are rather general, which means that in a large class of heterotic string models a non-perturbative 
superpotential cannot be generated. This raises a question of whether moduli in heterotic compactifications can ever be completely stabilized.  

The aim of this paper is to present  explicit examples where the non-perturbative superpotential is indeed non-zero. 


\subsection{Applicability of the residue theorem}
\label{assumption}


As  we have discussed, in the analysis of Beasley and Witten in~\cite{Beasley:2003fx} there is the assumption that  $\omega_{\td{X}} =\omega_{\cal{A}}|_{\td{X}}$. 
This assumption is necessary in order for their analysis to be a topological theory on the ambient space with $\tilde X$ as a vacuum solution--and, hence, to use their residue theorem.
It follows that, in their theorem, the area of all curves in~\eqref{1.8},~\eqref{new1} is measured using the Kahler form $\o_{{\cal A}}$
on ${\cal A}$ restricted to $\tilde X$. 
However, there are cases when this restriction is not the same as the physical Kahler form on $\tilde X$. 
Indeed, it is possible that  $h^{1,1} ({\tilde X})$ is not the same as $h^{1,1} ({\cal A})$ because there can be classes in $\tilde X$ 
which do not come as a restriction of classes from the ambient space. Hence, the residue theorem, strictly speaking, 
is valid only if $h^{1,1} ({\tilde X}) = h^{1,1} ({\cal A}).$\footnote{Such models were called favorable in~\cite{Anderson:2008uw}.} 
If   $h^{1,1} ({\tilde X}) > h^{1,1} ({\cal A})$ the residue theorem, though still valid in the topological theory, is not directly applicable to the 
physical heterotic string theory. In the former case, the area of holomorphic curves is measured using $\o_{{\cal A}}|_{\tilde X}$. 
But in the physical theory, it is measured using the actual Kahler form $\o_{{\tilde X}}$ on $\tilde X$. As a result, the curves which have the same 
area with respect to $\o_{{\cal A}}|_{\tilde X}$ might have different area with respect to $\o_{\tilde X}$ and, hence, might lie in different homology classes.
More precisely, if $h^{1,1} (\tilde X) > h^{1,1} ({\cal A})$ we have
\be 
\o_{\tilde X} = \o_{{\cal A}}|_{\tilde X} +\Delta \o_{\tilde X}\,, 
\label{asm1}
\ee
where $\Delta \o_{\tilde X}$ is the contribution to the Kahler form on $\tilde X$ from the $(1, 1)$ classes which do not come as
a restriction of classes from the ambient space. Then the actual area of a curve $C$ is given by 
\be 
\int_C \o_{\tilde X}  = \int_C (\o_{{\cal A}}|_{\tilde X} +\Delta \o_{\tilde X}) \geq \int_C \o_{{\cal A}}|_{\tilde X}\,. 
\label{asm2}
\ee
Two curves $C_1$ and $C_2$ which satisfy 
\be 
\int_{C_1} \o_{{\cal A}}|_{\tilde X} = \int_{C_2} \o_{{\cal A}}|_{\tilde X}
\label{asm3}
\ee
and appear to lie in the same homology class from the viewpoint of the residue theorem can actually have different area due to different  contributions
from $\Delta \o_{\tilde X}$ and can lie in different homology classes. 

To say it differently, if $h^{1,1} (\tilde X) > h^{1,1} ({\cal A})$ the correlation functions in the topological theory studied in~\cite{Beasley:2003fx}
do not coincide with correlation functions in the physical heterotic string theory on $\tilde X$. Hence, the cancellation in the residue theorem 
does not imply an analogous cancellation in the physical theory.
However, we can still apply the residue theorem 
to the physical theory. If in the physical theory we ignore $\Delta \o_{\tilde X}$ and measure the area of all curves using $ \o_{{\cal A}}|_{\tilde X}$ only,
then we should have the same cancellation as in the topological theory. 
Nevertheless, it is important to emphasize that now the cancellation happens among the 1-loop determinants of  the curves in {\it different} homology classes
but having the same area measured by $ \o_{{\cal A}}|_{\tilde X}$.
If we restore the actual area using the Kahler form $ \o_{\tilde X}$ on $\tilde X$, the contributions of 
these curves might no longer cancel each other 
because they might lie in different homology classes and have different area. That is, in the physical theory whether or not curves in a given homology class cancel each other 
cannot be directly deduced from the residue theorem. 
Below, we will give an example where the cancellation cannot happen simply because each curve is unique in its homology class.


\subsection{Discrete torsion}
\label{tor}

Our discussion so far  has been missing an important ingredient called  discrete torsion.
In general, for an arbitrary complex manifold, X, the second homology group with integer coefficients is of the form 
\be 
H_2 ({\rm X}, {\mathbb Z})=  {\mathbb Z}^k \oplus {\rm G}_{tor}\,, \quad k >0\,, 
\label{1.9}
\ee
where $ {\mathbb Z}^k$ is the free part and ${\rm G}_{tor}$ is a finite group called discrete torsion. For example, a discrete torsion factor of 
$H_2 ({\rm X}, {\mathbb Z})$ can arise
when X is a quotient of another Calabi-Yau manifold by a freely acting discrete isometry group $K$--as we will discuss below.
The existence of the torsion classes affects the $B$-field. Indeed, the $B$-field is an arbitrary closed 2-form $dB=0$. 
However, in general, it implies that the field strength $H = dB$ vanishes in $H^3 ({\rm X}, {\mathbb R})$ but 
not necessarily in $H^3 ({\rm X}, {\mathbb Z})$. In the later case the integral $\int_C B$ is not defined because the moduli 
space of the $B$-field is not connected. From the Universal Coefficient Theorem (see e.g.~\cite{Bott}) it follows that 
\be 
H_2 ({\rm X}, {\mathbb Z})_{tor} = H^3 ({\rm X}, {\mathbb Z})_{tor}\,. 
\label{uct}
\ee
This means that there is one-to-one correspondence between the torsion elements of $H_2 ({\rm X}, {\mathbb Z})$ 
and the number of the connected components of the moduli space of the $B$-field. 
These connected components can be labeled by the characters of the discrete group ${\rm G}_{tor}$.
Since the $B$-field is not continuous, we have to replace the exponential prefactor in~\eqref{1.1} 
with a more general map from $H^2 ({\rm X}, {\mathbb Z}) \to {\mathbb C}^*$~\cite{Aspinwall:1994uj}. While we will continue to denote this map by
 \be 
e^{-S_{cl}}: H_2 ({\rm X}, {\mathbb Z}) \to {\mathbb C}^*\, , 
\label{1.10}
\ee
it is no longer given by expression~\eqref{1.55}.  Specifically, it now depends on the discrete choice of the connected 
component of the moduli space of the $B$-fields.\footnote{A discrete
choice of a vacuum is quite common in heterotic compactifications on non-simply connected Calabi-Yau manifolds. Other discrete choices 
may involve a choice of  the equivariant structure of a vector bundle or a choice of  a Wilson line.}
Choosing a different connected component gives a different map~\eqref{1.10}. 
It is possible to describe the map~\eqref{1.10} more explicitly. Let us define the complexified Kahler form
\be 
\o_{{\mathbb C}}= \sum_{I=1}^{h^{1,1}} T^I \o_I =\sum_{I=1}^{h^{1,1}} \Big (\phi^I + i \frac{t^I}{2 \pi \a'}\Big) \o_I\,. 
\label{tor1}
\ee
Then the map~\eqref{1.55} can be understood as 
\be 
C \to {\rm exp}[ i \int_{C} \o_{{\mathbb C}}] = {\rm exp} [i \o_{{\mathbb C}} \cdot  C]\,, 
\label{tor2}
\ee
where, in the last step, we view $\o_{{\mathbb C}}$ as the Poincare dual 4-cycle and $\o_{{\mathbb C}} \cdot  C$ is the intersection 
of this 4-cycle with the curve $C$. 
However, $\o_{{\mathbb C}}$ defined in~\eqref{tor1} is Poincare dual only to an element of the {\it free} part of $H_4 ({\rm X}, {\mathbb Z})$.
Clearly, $\o_{{\mathbb C}}$ should also contain a torsion part.
Let ${\rm G}_{tor}$ have $r$ generators $\b_1, \dots, \b_r$. Then the complete expression for $\o_{{\mathbb C}}$ is given by
\be 
\o_{{\mathbb C}}= \sum_{I=1}^{h^{1,1}} T^I \o_I + \sum_{\a=1}^r s^{\a} \b_{\a}\, , 
\label{tor4}
\ee
where, slightly abusing notation, we continue to use the same symbol for the complexified Kahler form including torsion.
Since $\b_{\a}$ are torsion elements, it follows that for any $\a$ there is an integer $m_{\a}$ for which $m_{\a} \b_{\a}=0$. 
Hence, we obtain
\be 
{\rm exp} [i \o_{{\mathbb C}} \cdot  C] = e^{i \a_I(C) T^I} \prod_{\a=1}^r e^{i s^{\a} \b_{\a} (C)}\,, \qquad \b_{\a} (C)=\b_{\a} \cdot C\,.
\label{tor5}
\ee
Since $m_{\a} \b_{\a}=0$ and $C$ is arbitrary, it follows  that $\chi_{\a}= e^{i s^{\a}}$ is an $m_{\a}$-th root of unity. 
Hence, $s^{\a}$ can take only discrete values parametrizing the connected components of the moduli space of the $B$-field. 
It also follows that $\chi_{\a}$ is a character of ${\rm G}_{tor}$. We conclude that mapping~\eqref{1.55} now generalizes to 
\be 
C \to e^{i \a_I(C) T^I} \prod_{\a=1}^r \chi_{\a}^{\b_{\a} (C)}\,. 
\label{1.10.1}
\ee
The precise values of $\chi_{\a}$ depends on the choice of the torsion part of $B$; that is, on the choice of the connected component. 
Clearly, all curves in the same homology class of $H_2 ({\rm X}, {\mathbb Z})$ have the same value of $\b_{\a} (C)$ and, hence, 
pick up the same character-dependent factor in~\eqref{1.10.1}.

Let us now refine eq.~\eqref{1.8} in the presence of discrete torsion. Let $[C]$ be the homology class of the curve $C$ in $H_2 ({\rm X}, {\mathbb R})={\mathbb R}^k $. 
As we have just discussed, the curves in $[C]$ do not necessarily  lie in the same homology class in $H_2 ({\rm X}, {\mathbb Z})$ because 
they might belong to different torsion classes. Curves belonging to different torsion classes pick up different characters under the map~\eqref{1.10.1}.
Hence, equation~\eqref{1.8} is modified to become
\be 
W ([C])=e^{i \a_I (C) T^I} \sum_{i=1}^{n_{[C]}}
\frac{{\rm Pfaff}  ({\bar \pt}_{ {\rm V}_{C_i} (-1)})  }{[ {\rm det} {\bar \pt}_{{\cal O}_{C_i} (-1)}]^2} \prod_{\a=1}^r \chi_{\a}^{\b_{\a} (C_i)}\,, \qquad [C] \in  H_2 ({\rm X}, {\mathbb R})\,. 
\label{1.11}
\ee
To find the complete non-perturbative superpotential, we have to sum over all homology classes $[C] \in  H_2 ({\rm X}, {\mathbb R})$.
Later in the paper, we will analyze expression~\eqref{1.11} for a specific example. 


\section{The Schoen manifold and the prepotential}
\label{CY}


\subsection{The Schoen manifold}
\label{schoen}


Having presented the generic discussion above, we now proceed to calculate the non-perturbative superpotential for a specific Calabi-Yau threefold. 
This manifold, denoted by $X$, is the quotient of a simply connected, complete intersection Calabi-Yau threefold, $\tilde X$--chosen to be a specific Schoen manifold~\cite{Schoen}--with respect 
to its fixed-point free symmetry
group $K= {\mathbb Z}_3 \times {\mathbb Z}_3$. This Schoen threefold is defined as follows.
We construct $\td{X}$ as a compete intersection in the ambient space ${\cal A}= {\mathbb P}^1 \times {\mathbb P}^2 \times {\mathbb P}^2$ with 
homogeneous coordinates 
\be 
([t_0:t_1],  [x_0: x_1: x_2], [y_0: y_1: y_2]) \in {\mathbb P}^1 \times {\mathbb P}^2 \times {\mathbb P}^2\,. 
\label{2.1}
\ee
$\td{X}$ is then given by a common zero locus of two polynomial equations 
\bea
&&
P_1 (t_0, t_1) Q_1 (x_0, x_1, x_2) + P_2 (t_0, t_1) Q_2 (x_0, x_1, x_2) =0\,, \nonumber \\
&&
P_3 (t_0, t_1) Q_3 (y_0, y_1, y_2) + P_4 (t_0, t_1) Q_4 (y_0, y_1, y_2) =0\,.
\label{2.2}
\eea
Here $P_1, \dots, P_4$ are homogeneous polynomials of degree 1 and $Q_1, \dots, Q_4$ are homogeneous polynomials of degree 3. 
For the purposes of this paper, we will restrict $\td{X}$ to be given by the following polynomials 
\bea
&&
F_1= t_0 (x_0^3+ x_1^3 + x_2^3) + t_1 (x_0 x_1 x_2)=0 \,, \nonumber \\
&&
F_2 = (\l_1 t_0 +t_1)  (y_0^3+ y_1^3 + y_2^3) + (\lambda_2 t_0 + \l_3 t_1) (y_0 y_1 y_2)=0 \,. 
\label{2.3}
\eea
This manifold is self-mirror with $h^{1,1}= h^{2,1}=19$~\cite{Schoen, Ovrut:2002jk}.  Note that 
\be
h^{1, 1} (\td{X}) > h^{1, 1} ({\cal A})=3\, . 
\label{burt1}
\ee
It follows that on $\td{X}$ there are 16 $(1, 1)$ classes which do not arise as the restriction of $(1, 1)$ classes from the ambient space. 
The manifold $\td{X}$ defined by~\eqref{2.3} is invariant under the action of the $K= {\mathbb Z}_3 \times {\mathbb Z}_3$ symmetry generated by 
\begin{equation}
  \begin{split}
    g_1:&\;
    \begin{cases}
      [x_0:x_1:x_2] \mapsto
      [x_0:\zeta x_1:\zeta^2 x_2]
      \\
      [t_0:t_1] \mapsto
      [t_0:t_1] 
      ~\text{(no action)}
      \\
      [y_0:y_1:y_2] \mapsto
      [y_0:\zeta y_1:\zeta^2 y_2]
    \end{cases}
    \\
    g_2:&\;
    \begin{cases}
      [x_0:x_1:x_2] \mapsto
      [x_1:x_2:x_0]
      \\
      [t_0:t_1] \mapsto
      [t_0:t_1] 
      ~\text{(no action)}
      \\
      [y_0:y_1:y_2] \mapsto
      [y_1:y_2:y_0]\,,
    \end{cases}
  \end{split}
  \label{2.4}
\end{equation}
where $\zeta= e^{2 \pi i/3}$.
Note that this discrete symmetry does not act on ${\mathbb P}^1$. This action has fixed points on the ambient space ${\cal A}$, but 
not on $\td{X}$. Having constructed $\td{X}$ with a free ${\mathbb Z}_3 \times {\mathbb Z}_3$ action, we define
\be 
X =\td{X}/({\mathbb Z}_3 \times {\mathbb Z}_3)\,. 
\label{2.6}
\ee
This manifold is also self-mirror with $h^{1, 1}= h^{2,1 }=3$~\cite{Braun:2007xh}. From these Hodge numbers, it follows that 
$H_2 (X, {\mathbb R})= {\mathbb R}^3$. However, $H_2 (X, {\mathbb Z})$ is more involved. It was shown in~\cite{Braun:2007xh} 
that it contains the discrete torsion subgroup
\be
{\rm G}_{tor}=\mathbb{Z}_{3} \oplus \mathbb{Z}_{3} \, .
\label{burt2}
\ee
That is, the complete $H_2 (X, {\mathbb Z})$ is given by
\be 
H_2 (X, {\mathbb Z})=  {\mathbb Z}^3 \oplus {\mathbb Z}_3 \oplus {\mathbb Z}_3\,. 
\label{2.7}
\ee
Let us point out that from the Universal Coefficient Theorem it follows that 
\be 
H^2 (X, {\mathbb Z})_{tor}= H_1 (X, {\mathbb Z})_{tor} = \mathbb{Z}_{3} \oplus \mathbb{Z}_{3}\,. 
\label{uct1}
\ee
Hence, in the present case, the torsion groups of $H_2 (X, {\mathbb Z})$ and of $H^2 (X, {\mathbb Z})$ are the same. 

Let us present some mathematical details of $\td{X}$ and $X$ following~\cite{Braun:2007xh}. From eq.~\eqref{2.3} we see that 
for fixed $[t_0:t_1]$ we have two elliptic curves, one in each ${\mathbb P}^2$. Thus, each equation in~\eqref{2.3} defines a 
rational elliptic surface $dP_9 \in {\mathbb P}^1 \times {\mathbb P}^2$ and, hence,  $\td{X}$  is a double 
elliptic fibration over ${\mathbb P}^1$. The structure of $\td{X}$ can be illustrated using the diagram 
\begin{equation}
  \label{2.8}
  \vcenter{\xymatrix@!0@=12mm{
      \dim_\C=3: && & \td{X} \ar[dr]^{\pi_2} \ar[dl]_{\pi_1} \\
      \dim_\C=2: && B_1 \ar[dr] & & 
        B_2 \,. \ar[dl] \\
      \dim_\C=1: && & \CP^1
  }}
\end{equation}
Here $B_1$ and $B_2$ are the $dP_9$ surfaces given by the individual equations in~\eqref{2.3}.
The ${\mathbb Z}_3 \times {\mathbb Z}_3$ action descends to $B_1$ and $B_2$. Since its action is trivial on ${\mathbb P}^1$, 
on each $B_{k}$ for $k=1,2$ the ${\mathbb Z}_3 \times {\mathbb Z}_3$ must act by translation along the fiber by two independent sections of order 3. To simplify notation, we denote these sections on either $B_{k}$ by the same symbols $\mu$ and $\nu$--unless it is necessary to distinguish them. Additionally, each $B_k$ 
has the zero section $\sigma$. This determines the structure of Kodaira fibers and the Mordell-Weil group to be~\cite{Persson} 
\bea
&& 
{\rm sing} (B_1)= {\rm sing} (B_2) = 4 I_3\,, 
\nonumber \\
&& 
MW (B_1)= MW(B_2) = {\mathbb Z}_3 \oplus {\mathbb Z}_3\,. 
\label{2.9}
\eea
The Mordell-Weil group is generated by the zero section $\sigma$ and by the sections $\mu$ and $\nu$ of order 3. 
Each $B_k$ has 4 $I_3$ singular fibers, each containing 3 exceptional classes intersecting in a triangle. These classes will be denoted by 
$\theta_{ji}$, where $j=1, \dots, 4$ labels the singular fibers and $i=1, 2, 3$ labels the exceptional classes in each such fiber. As was shown in~\cite{ Shioda, OS},
the basis in $H_2 (B_k, {\mathbb Z})$ can be chosen to be 
\be 
H_2 (B_k, {\mathbb Z}) ={\rm span}_{{\mathbb Z}} \{\s, \mu, \nu, f, \theta_{11}, \theta_{21}, \theta_{31}, \theta_{32}, \theta_{41}, \theta_{42}
\}\,, 
\label{2.10}
\ee
where $f$ is the class of the elliptic fiber. The intersection numbers of these classes can be found in~\cite{Braun:2007xh}. Out of these basis elements, 
it is possible to construct divisors which are Poincare dual to  ${\mathbb Z}_3 \times {\mathbb Z}_3$ invariant $(1, 1)$ classes in $B_k$ (here we will use the same notation for divisors  and Poincare dual $(1, 1)$-forms).
The invariant cohomology group of $B_k$  is two-dimensional and generated by~\cite{Braun:2007xh}
\be 
H^2 (B_k, {\mathbb Z})_K = {\rm span}_{{\mathbb Z}} \{ f, t \}\,, 
\label{2.11}
\ee
where $t$ is a specific linear combination of the classes in~\eqref{2.10} given by
\be 
t= -3 \s - 3f + 3 \mu + 3\nu  + \theta_{11} +  \theta_{21} +2 \theta_{31} +2  \theta_{32} + 3 \theta_{41} + \theta_{42}\,. 
\label{2.12}
\ee
The intersection numbers of $f$ and $t$ are~\cite{Braun:2007xh}
\be 
f^2=0\,, \qquad f \cdot t =3\,, \qquad t^2 =1\,. 
\label{2.13}
\ee

Using the invariant cohomology classes in $B_k$, we can now construct divisors in $\td{X}$ Poincare dual to the invariant $(1, 1)$ classes on $\tilde{X}$. 
The invariant generators can be defined using the diagram~\eqref{2.8} and 
the invariant classes in~\eqref{2.11}: 
\bea
&&
\phi = \pi_1^{-1} (f_{1})= \pi_2^{-1} (f_{2})\,, 
\nonumber \\
&& 
\tau_1 = \pi_1^{-1} (t_{1})\,, \quad \tau_2 = \pi_2^{-1} (t_{2})\,. 
\label{2.15}
\eea
Let us now denote the corresponding Poincare dual $(1,1)$-forms as $\omega_{\phi}, \ \omega_{\tau_1}, \ \omega_{\tau_2}$. They form a basis of 
the invariant cohomology group 
$H^2 (\td{X}, \mathbb Z)_K$, and will descend to the quotient manifold $X$. 
Their triple intersection numbers can be found using the diagram~\eqref{2.8} and eq.~\eqref{2.13}
to be
\bea
&& 
\int_{\td{X}} \omega_{\phi} \wedge \omega_{\tau_1} \wedge \omega_{\tau_2} = \phi \cdot \t_1 \cdot \t_2= 9\,, 
\nonumber \\
&&
\int_{\td{X}} \omega_{\tau_1} \wedge \omega_{\tau_1} \wedge \omega_{\tau_2} = \t_1 \cdot \t_1 \cdot \t_2= 3\,, 
\nonumber \\
&&
\int_{\td{X}} \omega_{\t_1} \wedge \omega_{\tau_2} \wedge \omega_{\tau_2} = \t_1 \cdot \t_2 \cdot \t_2= 3\,.
\label{2.16}
\eea
The remaining triple intersection numbers are zero. We can now, somewhat abusing notation, define the  form $\o_X$ on $\td{X}$ by 
\be 
 \omega_X = t^1  \omega_{\phi}  + t^2 \omega_{\t_1} +  t^3 \omega_{\t_2}\,, \quad t^I>0  \ ,
\label{2.17}
\ee
which will descend to the non-torsion part of the Kahler form on $X$. 
Let us emphasize, however, that 
$\o_X$ is not the same as the Kahler form on $\tilde{X}$. This follows from the fact that  
on  $\tilde{X}$ there are additional classes of $H^2 (\td{X}, \mathbb Z)$ 
which are not invariant under ${\mathbb Z}_3 \times {\mathbb Z}_3$.
Indeed, as we stated previously, $h^{1, 1}( \td{X})=19$. This means that there are 16 $(1, 1)$ classes on $\td{X}$ in addition to $\omega_{\phi}$, $ \omega_{\tau_1}$ and $ \omega_{\tau_2}$. That is, the complete Kahler form on $\td{X}$ is given by 
\be 
\omega_{\td{X}} = \o_X + \Delta \o_{\td{X}}\,, 
\label{2.17.0}
\ee
where $ \Delta \o_{\td{X}}$ stands for the contribution from the additional 16 non-invariant classes. Comparing this expression with~\eqref{asm1}, we conclude that
\be
\o_X =  \o_{{\cal A}}|_{\td{X}} \ .
\label{burt3}
\ee
We will give an explanation for this relationship in the following subsection.


\subsection{The ambient space description of the invariant $(1, 1)$ classes}
\label{ambient}


The above description of the invariant $(1, 1)$ classes on $\td{X}$ is somewhat abstract. Here, we will give a simpler description of 
$\o_{\phi}, \o_{\t_1}, \o_{\t_2}$ in terms of the forms on the ambient space. Let ${\cal J}_1$,  ${\cal J}_2$,  ${\cal J}_3$ be the Kahler forms 
on the three projective spaces ${\mathbb P}^1$, ${\mathbb P}^2$, ${\mathbb P}^2$ forming the ambient space, normalized as
\be 
\int_{{\mathbb P}^1} {\cal J}_1 =1\,, \qquad \int_{{\mathbb P}^2} {\cal J}_2 \wedge {\cal J}_2 =1\,, \qquad
\int_{{\mathbb P}^2} {\cal J}_3 \wedge   {\cal J}_3 =1\,.
\label{2.18}
\ee
Since ${\mathbb Z}_3\times {\mathbb Z}_3$ does not act on ${\mathbb P}^1$, the cohomology class  of ${\cal J}_1$ is automatically invariant 
under ${\mathbb Z}_3\times {\mathbb Z}_3$. In the cohomology class of ${\cal J}_2$ (and similarly of ${\cal J}_3$), one can choose a representative to be  
the Kahler form of the Fubini-Study metric with the Kahler potential 
\be 
\log ( |x_0|^2 +|x_1|^2 +|x_2|^2 )\,, 
\label{2.18.1}
\ee
which is invariant under ${\mathbb Z}_3\times {\mathbb Z}_3$. This means that the cohomology classes of ${\cal J}_2$ and   ${\cal J}_3$  are also
invariant classes. 

Let us now define the $(1, 1)$ classes on $\td{X}$ by restriction
\be
J_1={\cal J}_1|_{\td{X}}\,, \qquad J_2={\cal J}_2|_{\td{X}}\,, \qquad J_3={\cal J}_3|_{\td{X}}\,. 
\label{2.19}
\ee
By construction, the cohomology classes of $J_1, J_2, J_3$ are invariant classes in $H^2 (\td{X}, {\mathbb Z})$ and,  hence, form a basis
in $H^2 (\td{X}, {\mathbb Z})_K$. The triple intersection numbers of $J_1, J_2, J_3$ can be computed by the standard methods of complete intersection 
Calabi-Yau manifolds (see e.g.~\cite{Hubsch})\footnote{One can also compute them by lifting the triple intersection integrals to the ambient space by inserting
the delta-function current as in~\cite{Candelas:1987se, Blesneag:2015pvz, Blesneag:2016yag}. 
See Appendix A in~\cite{Blesneag:2015pvz} for a similar calculation.}  with the following result
\bea
&& 
\int_{\td{X}} J_1 \wedge J_2 \wedge J_3 =9\,, 
\nonumber \\
&& 
\int_{\td{X}} J_2 \wedge J_2 \wedge J_3 =3\,,  \qquad 
\int_{\td{X}}  J_2 \wedge J_3 \wedge J_3 =3\,, 
\label{2.20}
\eea
with the remaining ones being zero. Comparing eq.~\eqref{2.20} with~\eqref{2.16} we conclude that 
\be 
\o_{\phi}= J_1\,, \qquad \o_{\t_1}= J_2\,, \qquad \o_{\t_2}= J_3\,.
\label{2.21}
\ee
That is, the invariant $(1, 1)$ classes $\o_{\phi}, \o_{\t_1}, \o_{\t_2}$,  which were constructed in the previous subsection in a rather abstract 
way, are simply the restriction of the Kahler forms on the projective spaces forming the ambient space--thus explaining expression~\eqref{burt3}.
Due to the normalization properties~\eqref{2.18},  the forms $J_1, J_2, J_3$ can be viewed as first Chern classes of the following line bundles on ${\cal A}$:
\be 
{\cal O}_{{\cal A}} (1, 0, 0)\,, \qquad {\cal O}_{{\cal A}} (0, 1, 0)\,, \qquad {\cal O}_{{\cal A}} (0, 0, 1)\,. 
\label{2.21.1}
\ee
This implies the following  relations between the line bundles 
\be
{\cal O}_{\td{X}} (\phi) = {\cal O}_{{\cal A}} (1, 0, 0)|_{\td{X}}\,, \quad 
{\cal O}_{\td{X}} (\t_1) = {\cal O}_{{\cal A}} (0, 1, 0)|_{\td{X}}\,, \quad 
{\cal O}_{\td{X}} (\t_2) = {\cal O}_{{\cal A}} (0, 0, 1)|_{\td{X}}\,.
\label{2.22}
\ee
These relations will be useful later. For emphasis, we again note note that the Kahler form~\eqref{2.17.0} can be written as 
\be 
\omega_{\td{X}} = \o_{{\cal A}}|_{\td{X}} + \Delta \o_{\td{X}}\,.
\label{2.21.2}
\ee
We see, therefore, that the Kahler form on $\td{X}$ is not simply given by the restriction of the Kahler form from the ambient space but, rather,  contains
an additional term $ \Delta \o_{\td{X}}$. 


\subsection{The prepotential and Gromov-Witten invariants}
\label{prepot}


The number of holomorpic, isolated, genus 0  curves in each homology class of $X$ can be read off 
from the prepotential in type II string theory. The prepotential on $X$ was
computed in~\cite{Braun:2007xh, Braun:2007tp, Braun:2007vy}. In this subsection, we will review the result. 
Since the $(1, 1)$ classes  $\{\o_{\phi}, \o_{\t_1}, \o_{\t_2}\}$ are ${\mathbb{Z}}_{3} \times {\mathbb{Z}}_{3}$ invariant on $\td{X}$,  they descend to cohomology classes on $X$. To 
simplify our notation,  we will label these cohomology classes using the same symbols. Let $\{ [C_{\phi}],  [C_{\t_1}], [C_{\t_2}]\}$ be the dual homology classes in the 
free part of $H_2 (X, {\mathbb Z})$. If  $C_{\phi}$, $C_{\t_1}$, $C_{\t_2}$ are arbitrary representatives of these classes, then
\be 
\int_{C_{\phi}} \o_{\phi} =1\,, \qquad \int_{C_{\t_1} }\o_{\t_1} =1\,, \qquad \int_{C_{\t_2}} \o_{\t_2} =1 \ ,
\label{2.17.1}
\ee
with the other integrals being zero.
Let us define
\bea
&&
p= e^{-S_{cl} } ( [C_{\phi}]) =  {\rm exp} \Big[  \int_{C_{\phi}} \Big( -\frac{\o_X}{2\pi\a'} + i B\Big)\Big] = e^{i T^1}\,, 
\nonumber \\
&&
q= e^{-S_{cl} } ( [C_{\t_1}]) ={\rm exp} \Big[  \int_{C_{\t_1}} \Big( -\frac{\o_X}{2\pi\a'} + i B\Big)\Big] = e^{i T^2}\,, 
\nonumber \\
&&
r=e^{-S_{cl} } ( [C_{\t_2}]) = {\rm exp} \Big[  \int_{C_{\t_2}} \Big( -\frac{\o_X}{2\pi\a'} + i B\Big)\Big] = e^{i T^3}\,.
\label{2.17.2}
\eea
Since $H_2 (X, {\mathbb Z})$ contains  torsion classes, we also have to introduce the image under the map $e^{-S_{cl}}$ of the ${\mathbb{Z}}_{3} \times {\mathbb{Z}}_{3}$ torsion generators--which we denote by $b_1$ and $b_2$ respectively and satisfy $b_1^3= b_2^3=1$. 
Let $[C]$ be a homology class of the form
\be 
[C]= (n_1, n_2, n_3, m_1, m_2) \in H_2 (X, {\mathbb Z}) = {\mathbb Z}^3 \oplus {\mathbb Z}_3 \oplus {\mathbb Z}_3\,, \qquad  m_1, m_2 =0, 1, 2\,.
\label{2.17.4}
\ee
Then, as shown in ~\eqref{1.10.1}, the image of this class under $e^{-S_{cl}}$ is given by 
\be 
e^{-S_{cl} } ( [C]) = p^{n_1} q^{n_2} r^{n_3} b_1^{m_1} b_2^{m_2}\,,
\label{2.17.6}
\ee
where $b_1^{m_1} b_2^{m_2}$ is a ${\mathbb Z}_3 \times  {\mathbb Z}_3$  character.
The prepotential in type II string theory is defined by the expression 
\be 
{\cal F}_X = \sum_{[C] \in H_2 (X, {\mathbb Z})} n_{[C]} {\rm Li}_3 (e^{-S_{cl} } ( [C]) ) =
 \sum_{[C] \in H_2 (X, {\mathbb Z})} n_{[C]} {\rm Li}_3 
(p^{n_1} q^{n_2} r^{n_3} b_1^{m_1} b_2^{m_2})\,.
\label{2.17.7}
\ee
Here the sum is over all holomorphic, isolated, genus 0 curves and 
the polylogarithm ${\rm Li}_3$ takes proper care of multiple wrappings. If we know ${\cal F}_X $, 
we can expand it in powers of $p, q, r, b_1, b_2$ and read off the Gromov-Witten invariants $n_{[C]}$. 
The prepotential ${\cal F}_X$ for the quotient Calabi-Yau manifold $X$ in~\eqref{2.6} was computed in~\cite{Braun:2007xh, Braun:2007tp, Braun:2007vy}.
Here, we present the result to low orders in $p, q, r$. It is given by
\be 
{\cal F}_X = p (1+ b_1+ b_1^2) (1+ b_2+ b_2^2) {\cal P} (q)^4  {\cal P} (r)^4 + {\cal O} (p^2)\,, 
\label{2.17.8}
\ee
where the polynomial ${\cal P}(q)$ is of the form 
\be
{\cal P}(q) = 1+ {\cal O} (q)\,.
\label{2.17.9}
\ee

Let us discuss some simple consequences of eqs.~\eqref{2.17.8}, \eqref{2.17.9}. It follows that there are no terms $\sim p^0$. 
In other words, there are no isolated, genus 0 curves in the homology 
classes $(0, n_2, n_3, m_1, m_2)$. However, there are terms in ${\cal F}_X$ that are $\sim p^{1}$. Hence, there are isolated, genus zero 0 curves in the $(1, n_2, n_3, m_1, m_2)$ homology classes. It follows from~\eqref{2.17.2} that the contribution of these classes to $e^{-S_{cl}}$ and, hence, the superpotential is proportional to 
$e^{i T^1 +i n_2 T^2 + i n_3 T^3}$. This means that the leading contribution to the superpotential is $\sim e^{i T^1}$; all terms with $\ n_2 > 0$ and/or $\ n_3 > 0$ being exponentially suppressed. Similarly, the contribution to the superpotential of any class with $n_{1} >1$ is also suppressed relative to $ e^{i T^1}$. Therefore, since we are interested in computing the superpotential, we will focus on the homology classes of the form $(1, 0, 0, m_1, m_2)$.
The number of  isolated, genus 0 curves in each such class can be read off from the most leading term in~\eqref{2.17.8}. This is given by 
\be 
{\cal F}_X  \sim p (1+ b_1+ b_1^2) (1+ b_2+ b_2^2) \,. 
\label{2.17.10}
\ee
%
It follows that in each torsion class there is precisely 1 curve. That is, 
\be 
n_{[C]}=1 ~~~{\rm for~each~class} ~~~  [C]= (1, 0, 0, m_1, m_2)\,, ~  m_1, m_2 =0, 1, 2\,. 
\label{2.17.11}
\ee
The main aim of the rest of this paper will be to compute the non-perturbative superpotential~\eqref{1.11} summed over these 9 isolated, genus 0 curves. All of them 
are in the same homology class in $H_2 (X, {\mathbb R})$ and, hence, have the same area with respect to 
the Kahler form on $X$. However, they are distributed in 9 different homology classes once we take discrete torsion into account. 


\subsection{Explicit construction of the isolated, genus 0 curves}
\label{curves}


It is possible to explicitly  visualize these 9  curves as follows.
We see from eq.~\eqref{2.17.2} that, ignoring torsion, they are in the same homology class $[C_{\phi}]$ which is dual to 
the class of the $(1, 1)$-form $\omega_{\phi}$. Let us lift these curves from $X$ to  $\td{X}$. 
Since on $\td{X}$  we have $\omega_{\phi} = J_1={\cal J}_1|_{\td{X}}$, where ${\cal J}_1$ is the Kahler 
form on ${\mathbb P}^1 \subset {\cal A}$, the pre-image of these 9 curves gives 81 holomorphic, genus 0 curves 
on $\td{X}$ which can be parametrized by $[t_0: t_1]$.
Hence, we can visualize these curves by demanding that eqs.~\eqref{2.3} are solved 
for arbitrary $[t_0:t_1]$. This is equivalent to solving the system of equations 
 \be 
 x_0 x_1 x_2=0\,, \quad x_0^3+ x_1^3+ x_2^3 =0\,, \quad  y_0 y_1 y_2=0\,, \quad y_0^3+ y_1^3+ y_2^3 =0
 \label{2.23}
 \ee
on ${\mathbb P}^2 \times {\mathbb P}^2 $. It is easy to see that this system is solved by  $9 \times 9=81$ distinct points
on ${\mathbb P}^2 \times {\mathbb P}^2 $. Since the solutions of~\eqref{2.23} are distinct points, all the corresponding curves
in $\td{X}$ are isolated. In Appendix~\ref{A} we compute the normal bundle for each of these curves and check that it is indeed ${\cal O}(-1) \oplus  {\cal O}(-1)$.
Due to the ${\mathbb Z}_3 \times {\mathbb Z}_3$ symmetry, the above 81 curves split into 9 orbits under the action of ${\mathbb Z}_3 \times {\mathbb Z}_3$--each
orbit containing 9 curves. 
The curves in the same orbit are obtained from each other by the action of the ${\mathbb Z}_3 \times {\mathbb Z}_3$ group. When we descend
to the quotient manifold $X$, all curves in one orbit yield the same curve in $X$. Hence, we obtain 9 isolated, genus 0 curves 
in $X$ which are precisely the curves discussed at the end of the previous subsection. It follows from the prepotential~\eqref{2.17.10} 
they are in the same homology class in 
$H_2 (X, {\mathbb R})$ but  in 9 different homology classes in $H_2 (X, {\mathbb Z})$.

We now present 9 curves in $\td{X}$ which do not lie in the ${\mathbb Z}_3 \times {\mathbb Z}_3$ orbits of each other and 
which, therefore, descend to 9 distinct curves in $X$. To accomplish this, let us write the generators $g_1$ and $g_2$ in~\eqref{2.4} in the matrix form 
\begin{eqnarray}
g_1 = \begin{pmatrix} 1& 0 & 0  \\ 0 & \xi & 0 \\  0 & 0 &  \xi^2    \end{pmatrix} \,, \quad
g_2 = \begin{pmatrix} 0& 1 & 0  \\ 0 & 0 & 1 \\  1 & 0 &  0    \end{pmatrix} \,.
\label{2.24}
\end{eqnarray}
Since ${\mathbb Z}_3 \times {\mathbb Z}_3$ acts simultaneously on both ${\mathbb P}^2$'s,
it is convenient to combine  $[x_0: x_1: x_2]$ and  $[y_0: y_1: y_2]$ into a 6-vector $(x_0, x_1, x_2, y_0, y_1,  y_2)^T$. In this basis, the generators of 
${\mathbb Z}_3 \times {\mathbb Z}_3$ are 
\begin{eqnarray}
K_1 = \begin{pmatrix} g_1 & 0   \\ 0 & g_1   \end{pmatrix} \,, \quad
K_2 = \begin{pmatrix} g_2 & 0   \\ 0 & g_2     \end{pmatrix} \,.
\label{2.25}
\end{eqnarray}
Now choose one arbitrary solution of~\eqref{2.23}. For example, pick 
\be 
s_1= (1, -1, 0, 1, -1, 0)^T \ ,
\label{2.27}
\ee
where the symbol ``T'' means think of this as a column vector.
It corresponds to the curve 
\be 
C_1 ={\mathbb P}^1 \times s_1 = 
[t_0 :t_1]  \times [1: -1 :0] \times [1: -1 :0] 
\subset \td{X} \subset {\mathbb P}^1 \times {\mathbb P}^2 \times {\mathbb P}^2\,. 
\label{2.27.1}
\ee
Let us now construct the remaining 8 curves $C_i =  {\mathbb P}^1 \times s_i$, $i=2,\dots ,9$  by acting on $s_{1}$ as follows:
\begin{eqnarray}
&&
s_2 = \begin{pmatrix} g_1 & 0   \\ 0 & 1   \end{pmatrix} s_1 \,, \quad
s_3 = \begin{pmatrix} 1 & 0   \\ 0 & g_1     \end{pmatrix} s_1 \,, \quad 
s_4 = \begin{pmatrix} g_2 & 0   \\ 0 & 1   \end{pmatrix} s_1 \,, \quad
s_5 = \begin{pmatrix} 1 & 0   \\ 0 & g_2     \end{pmatrix} s_1 \,, \nonumber \\
&&
s_6 = \begin{pmatrix} g_1 g_2 & 0   \\ 0 & 1   \end{pmatrix} s_1 \,, \quad
s_7 = \begin{pmatrix} 1 & 0   \\ 0 & g_1 g_2     \end{pmatrix} s_1 \,, \quad 
s_8 = \begin{pmatrix} g_1 & 0   \\ 0 & g_2   \end{pmatrix} s_1 \,, \quad
s_9 = \begin{pmatrix} g_2& 0   \\ 0 & g_1     \end{pmatrix} s_1 \,.
\label{2.28}
\end{eqnarray}
One can check that these curves solve eqs.~\eqref{2.23} and
cannot be obtained from each other by the action of ${\mathbb Z}_3 \times {\mathbb Z}_3$. 


\section{Non-vanishing of the superpotential on $\td{X}$}


Let us now compactify the $E_{8} \times E_{8}$ heterotic string on the manifold $\td{X}$, and consider the non-perturbative superpotential 
generated by the 81 isolated, genus 0  curves discussed above.
To describe the complete string vacuum, one must also introduce a specific holomorphic vector 
bundle--which we will do in the next section. Although we will compute the superpotential for this specific bundle, the results of the present section are valid for any vector bundle.  
Since the curves specified by the solutions of~\eqref{2.23} are all parameterized by $[t_0: t_1]$, 
it is tempting to conclude that they are in the same homology class dual to the $(1, 1)$ class $\o_{\phi}={\cal J}_1|_{\td{X}}$.
However, we will see that this is not the case.  In fact, we will show that these curves lie in 81 different homology classes. 

First consider the twofold $B_1 \simeq dP_9 \subset {\mathbb{P}} \times {\mathbb{P}}^{2}$ defined by 
\be
F_1= t_0 (x_0^3+ x_1^3 + x_2^3) + t_1 (x_0 x_1 x_2)=0 \ .
\label{supp1}
\ee
Let us now examine the genus 0 curves parametrized by $[t_0: t_1]$. 
They are specified by the 9 solutions of
 \be 
 x_0 x_1 x_2=0\,, \quad x_0^3+ x_1^3+ x_2^3 =0 \ .
 \label{supp2}
 \ee
Each solution is a distinct section  of the elliptically fibered surface $B_1 \simeq dP_9$. 
We denote these sections by $\s_i$, $i=1, \dots, 9$. 
Since the order of the Mordell-Weil group $MW (B_1)= Z_3 \oplus Z_3 $ is 9, these sections are 
 in one-to-one correspondence with the elements of $MW (B_1)$~\cite{Braun:2007xh}.
On the other hand, it was shown in~\cite{Shioda} that on a $dP_9$ twofold distinct elements of the Mordell-Weil group
are all non-homologous to each other. It follows that the 9 solutions to~\eqref{supp2} fall into 9 different homology classes. 
The same is true on $B_2 \simeq dP_{9}$ defined by
\be
F_2 = (\l_1 t_0 +t_1)  (y_0^3+ y_1^3 + y_2^3) + (\lambda_2 t_0 + \l_3 t_1) (y_0 y_1 y_2)=0 
\label{burt4}
\ee
 for  the 9 genus 0 curves specified by the solutions of
 \be
 y_0 y_1 y_2=0\,, \quad y_0^3+ y_1^3+ y_2^3 =0 \ . 
 \label{burt5}
 \ee
 For reasons of simplicity, we also denote these sections by $\s_i$, $i=1, \dots, 9$.
 
We now extend this result to $\td{X}$.
To start with, note that there is a natural map 
\be 
H_2 (\td{X}, {\mathbb Z}) \to  H_2 (B_1, {\mathbb Z}) \times  H_2 (B_2, {\mathbb Z})\,. 
\label{map1}
\ee
Let us define 
\be 
\s_i \underline{\times} \s_j = (\s_i \times \s_j) \cap \td{X}  \in H_2 (\td{X}, {\mathbb Z}) \,, \qquad \s_i \in MW (B_1)\,, 
\quad \s_j \in MW (B_2)\,.
\label{map2}
\ee
This provides a more abstract way to visualize the 81 curves solving eqs.~\eqref{2.23}. 
Then  the map~\eqref{map1}  acts on the above described solutions as 
\be
\s_i \underline{\times} \s_j  \to \s_i \times \s_j \in H_2 (B_1, {\mathbb Z}) \times  H_2 (B_2, {\mathbb Z})\,.
\label{map3}
\ee
We note that the 81 elements $\s_i \times \s_j$, all being distinct, are in one-to-one correspondence with elements of the Mordell-Weil group 
$MW (B_1 \times B_2)$  of $B_1 \times B_2$--which is given by $MW (B_1) \oplus MW (B_2)$ and is of order 81 as well. 
Therefore, we have a map between the set $\{\s_i \underline{\times} \s_j\}$ and the elements of $MW (B_1) \oplus MW (B_2)$. 
By construction, it is a surjective linear map between two finite sets consisting of 81 elements each. Hence, it is one-to-one. 
Since all distinct elements of $MW (B_1) \oplus MW (B_2)$ are non-homologous to each other, it follows that all 81 curves obtained in~\eqref{2.23}
lie in 81 different homology classes. In particular, it follows that each of these 81 homology classes has precisely 1 isolated, genus 0 curve. 
Hence, as long  as the Pfaffian of the Dirac operator of at least one of these curves is not identically zero--which is expected to be true 
for a generic vector bundle--the non-perturbative superpotential in this theory is non-zero. 
We will show this explicitly for a specific holomorphic vector bundle in the remainder of this paper.

To finish this section, let us point out that any of the 81 homology classes  discussed above are dual to a $(1, 1)$ class of the 
form ${\cal J}_1|_{\td{X}} +\Delta \omega_{\td{X}}$.\footnote{Explicit calculation of $\Delta \omega_{\td{X}}$ for each of the 81 curves 
is a tedious, complicated task which is unnecessary for the purposes of the paper.}
Here, it suffices to recall that $\Delta \omega_{\td{X}}$ stands for the classes on $\td{X}$ which cannot be obtained as a restriction from the ambient space. 
All of these 81 curves have equal area with respect to ${\cal J}_1|_{\td{X}}$ and 
$\o_{{\cal A}}|_{\td{X}}$, 
but have different areas with the respect to the actual Kahler form  $\o_{\td{X}}$. 
Hence, we have an explicit realization of the situation described in Subsection~\ref{assumption}. As was discussed, this violates 
one of the assumptions of the Beasley-Witten residue theorem and, hence, one can expect a non-vanishing instanton superpotential. Finally, note that since $\Delta \omega_{\td{X}}$ contains only non-invariant
classes under ${\mathbb Z}_3 \times {\mathbb Z}_3$ which vanish on the quotient manifold  $X$,
the area of the images of the 9 curves in eqs.~\eqref{2.27}, \eqref{2.27.1}, \eqref{2.28} in $X$ is the same with respect to the Kahler form $\o_X$ on $X$. 
That is, although the original 81 curves were in different homology classes on the covering manifold,
their  images are in the same homology class on the quotient manifold modulo torsion. 


\section{The vector bundle}
\label{bundle}


The rest of the paper will be devoted to an explicit computation of the superpotential in a  concrete example 
of a  theory on $X$. Specifically, we will consider 
a toy model where the holomorphic vector bundle $V$ is taken to have structure group $SU(3)$. 
A more phenomenologically realistic class of bundles will be studied elsewhere. 
Since $X$ is a quotient manifold, it is easiest to first construct a vector bundle $\td{V}$ on the covering space $\td{X}$
which is equivariant under the action of $K={\mathbb Z}_3 \times {\mathbb Z}_3$. The moduli space of an equivariant vector bundle $\td{V}$ 
consists of connected components labeled by the characters of $K$. The vector bundle $V$ on the quotient space $X$ is then defined as
\be 
V= \td{V}/({\mathbb Z}_3 \times {\mathbb Z}_3) \ ,
\label{3.1}
\ee
with the moduli space of $V$ consisting of one of the  connected components.
The choice of the connected component is referred to as to 
the choice of the equivariant structure. Different choices of equivariant structure give  different vector bundles $V$ in~\eqref{3.1}. 

First, we will discuss a construction of $\td{V}$ in terms of the data on $\td{X}$. We then will express the same vector bundle as
the restriction of  a vector bundle $\td{\cal V}$ on the ambient space. The second description is more explicit and will be used in computing 
the superpotential in the next section.


\subsection{Construction of the vector bundle $\td{V}$ on $\td{X}$}
\label{vonx}


We will construct $\td{V}$ by specifying the line bundles $L_1,  L_2, L_3$ on $\td{X}$ satisfying the property
\be 
L_1 \otimes L_2 \otimes L_3 ={\cal O}_{\td{X}} \,. 
\label{3.2}
\ee
Then we define $\td{V}$ as a sequence of  extensions
\bea
&& 
0 \longrightarrow L_1 \longrightarrow \td{W}  \longrightarrow L_2 \longrightarrow 0\,, 
\nonumber \\
&&
0 \longrightarrow \td{W} \longrightarrow \td{V}  \longrightarrow L_3 \longrightarrow 0\,. 
\label{3.3}
\eea
Eq.~\eqref{3.2} assures that the structure group of $\td{V}$ is $SU(3)$ rather than $U(3)$. The structure group of the rank 2 vector bundle $\td{W}$ is $U(2)$. 
For $\td{V}$ to descend to the quotient manifold $X$, it has to be equivariant.  To achieve that, it is sufficient to require that the line bundles 
$L_1, L_2, L_3$ are equivariant. A discussion of equivariant line bundles on the Schoen manifold can be found in~\cite{Braun:2005zv}. 
The action of the discrete group in~\cite{Braun:2005zv} was chosen to be different from ours in~\eqref{2.4}. However, the conclusions on equivariance are the same. 
Here we will simply state the conclusions, referring to~\cite{Braun:2005zv} for additional details. 

First, any equivariant line bundle $L$ on $\td{X}$ has to be constructed out of the invariant divisors in~\eqref{2.15}. That is, it has to be of the form 
\be 
L= {\cal O}_{\td{X}} (c_1 \phi + c_2 \tau_1+ c_3 \tau_2)\, ,
\label{3.4}
\ee
where $c_1, c_2, c_3$ are integers. 
In addition, the sum $c_2 +c_3$ has to be divisible by 3. 
In our toy model, we will choose $L_1, L_2, L_3$ to be 
\bea
&&
L_1= {\cal O}_{\td{X}} (-2 \phi + 2  \tau_1+  \tau_2)\,, \nonumber \\
&&
L_2= {\cal O}_{\td{X}} (  \tau_1 -  \tau_2)\,, \nonumber \\
&&
L_3= {\cal O}_{\td{X}} (2 \phi -3  \tau_1)\,.
\label{3.5}
\eea
Note that eq.~\eqref{3.2} is satisfied. We will take the trivial choice of the equivariant structure; that is, we will assume 
that the moduli space of $V$ given by~\eqref{3.1} consists of the component of the moduli space of $\td{V}$ 
which is invariant under ${\mathbb Z}_3 \times {\mathbb Z}_3$. 

For $V$ to have structure group $SU(3)$ rather than its  subgroup, we have to make sure that there exist non-trivial extensions in~\eqref{3.3}. 
The spaces of  non-trivial extensions are given by 
\be 
H^1 (\td{X},  L_1 \otimes L_2^*)  \qquad {\rm and} \qquad  H^1 (\td{X},  \td{W} \otimes L_3^*)
\label{3.6}
\ee
respectively. For simplicity, we will often denote them by $[\td{W}]$ and $[\td{V}_{\td{W}}]$.
Note that $H^1 (\td{X},  \td{W} \otimes L_3^*)$ is the space of extensions for a fixed extension $\td{W}$ in $[\td{W}]$. 
That is why we denote it by $[\td{V}_{\td{W}}]$. 
Each element in the extension class defines a vector bundle. However,
it is important to take into account that different elements in the extension class can define isomorphic vector bundles. 
Let $\td{W}_1$ and $\td{W}_2$ be two vector bundles from the same extension class $[\td{W}]$. That is, they both satisfy
\bea
&& 
0 \longrightarrow L_1 \longrightarrow \td{W}_1  \longrightarrow L_2 \longrightarrow 0\,, 
\nonumber \\
&&
0 \longrightarrow L_1 \longrightarrow \td{W}_2  \longrightarrow L_2 \longrightarrow 0\,. 
\label{3.6.1}
\eea
For any line bundle $L$ there is an  isomorphism $L \to \l L$, where we multiply all elements of the fiber of $L$
by a non-zero complex number $\l$.
Let us consider the following isomorphisms of $L_1$ and $L_2$: \  $L_1 \to L_1$, $L_2 \to \lambda L_2$.
Then from the ``five'' lemma (see e.g.~\cite{Bott}), 
it  follows that $W_1$ and $W_2$ are isomorphic. This means that elements in $H^1 (\td{X},  L_1 \otimes L_2^*) $
related by a multiplication by $\l \in {\mathbb C}^*$ correspond to  the same vector bundle.  
Therefore, the moduli space ${\cal M}({\td{W}})$ 
of vector bundles corresponding to the extension class $[\td{W}]$ is the projectivization of $H^1 (\td{X},  L_1 \otimes L_2^*)$:
\be 
{\cal M}({\td{W}}) ={\mathbb P} H^1 (\td{X},  L_1 \otimes L_2^*)\,.
\label{3.6.2}
\ee
Note that 
\be 
{\rm dim} {\cal M}({\td{W}} ) = h^1 (\td{X},  L_1 \otimes L_2^*) -1\,. 
\label{3.6.3}
\ee
Similarly, the moduli space ${\cal M}({\td{V}_{\td{W}}})$ of vector bundles  corresponding to the extension 
class $[\td{V}_{\td{W}}]$ is given by 
\be
{\cal M}({\td{V}_{\td{W}}})=  {\mathbb P} H^1 (\td{X},  \td{W} \otimes L_3^*)\,, \qquad {\rm dim} {\cal M}({\td{V}_{\td{W}}} )= h^1 (\td{X},  \td{W} \otimes L_3^*)-1\,. 
\label{3.7}
\ee
The full moduli space ${\cal M}({\td{V}})$ 
of $\td{V}$ can be then understood as a fibration over ${\cal M}({\td{W}}) ={\mathbb P} H^1 (\td{X},  L_1 \otimes L_2^*)$,  where the fiber 
at a fixed extension $\td{W}$ is given by ${\cal M}({\td{V}_{\td{W}}}) = {\mathbb P} H^1 (\td{X},  \td{W} \otimes L_3^*)$. 
In Appendix~\ref{B}, we compute the dimensions of the spaces in~\eqref{3.6}. We find
\be 
h^1 (\td{X},  L_1 \otimes L_2^*)= 18\,, \qquad   h^1 (\td{X},  \td{W} \otimes L_3^*)=117\,. 
\label{3.8}
\ee
This means that
\be 
{\rm dim} {\cal M}({\td{W}}) =17\,, \quad   {\rm dim}  {\cal M}({\td{V}_{\td{W}}}) =116
\label{3.9}
\ee
and, hence, 
\be
{\rm dim} {\cal M}({\td{V}}) =17+116=133\,. 
\label{burt6}
\ee
Note that if we introduce coordinates in the vector space of extensions, it is straightforward to introduce coordinates on its projectivization; that is,
we simply treat these coordinates as homogeneous ones. 

Since the line bundles $L_1, L_2, L_3$ are equivariant, they descend to the quotient manifold $X$. To simplify notation, we will denote the corresponding 
line bundles on $X$ by the same letters $L_1, L_2, L_3$. Hence, the vector bundles $W$ and $V$ on $X$, obtained by modding out $\td{W}$ and 
$\td{V}$ by the action of ${\mathbb Z}_3 \times {\mathbb Z}_3$, can be defined by the similar extension sequences on $X$
\bea
&& 
0 \longrightarrow L_1 \longrightarrow W  \longrightarrow L_2 \longrightarrow 0\,, 
\nonumber \\
&&
0 \longrightarrow W \longrightarrow V  \longrightarrow L_3 \longrightarrow 0\,. 
\label{3.10}
\eea
As we mentioned before, we will take the trivial choice of the equivariant structure. This means that the 
the extension classes $[W]$ and $[V_W]$ can be taken to be the invariant components of~\eqref{3.6}.
That is, 
\be
h^1 (X,  L_1 \otimes L_2^*)= 2\,, \qquad   h^1 (X,  W \otimes L_3^*)=13\,, 
\label{3.10.1}
\ee
where we have simply divided the dimensions in~\eqref{3.8} by the order of ${\mathbb Z}_3 \times {\mathbb Z}_3$. Then it follows that
\be 
{\rm dim} {\cal M}(W) =1\,, \quad   {\rm dim} {\cal M}(V_W)  =12\,, \quad   {\rm dim} {\cal M}(V) =1+12=13\,.
\label{3.11}
\ee
To show that $\td{V}$ and, hence, $V$ admits an Hermitian connection satisfying eq.~\eqref{1.5.1}, we need to prove that the extensions described above
correspond to  stable vector bundles. This is discussed in Appendix~\ref{C}. 


\subsection{The ambient space description of $\td{V}$}
\label{vona}


As was shown in the previous section, the line bundles ${\cal O}_{\td{X}} (\phi), {\cal O}_{\td{X}} (\t_1), {\cal O}_{\td{X}} (\t_2)$
can be obtained as restrictions of line bundles on the ambient space. Using~\eqref{2.22}, we find that $L_1, L_2, L_3$ 
are also restrictions of line bundles on ${\cal A}$. Let us define 
\be
{\cal L}_1= {\cal O}_{{\cal A}} (-2, 2, 1)\,, \quad {\cal L}_2= {\cal O}_{{\cal A}} (0, 1, -1)\,, \quad {\cal L}_3= {\cal O}_{{\cal A}} (2, -3, 0)\,. 
\label{3.12}
\ee
Then $L_1= {\cal L}_1|_{\td{X}}, L_2= {\cal L}_2|_{\td{X}}, L_3= {\cal L}_3|_{\td{X}}$. This implies that the extensions $\td{W}$ and $\td{V}$ 
are also restrictions of extensions on ${\cal A}$, which we denote by  $\td{{\cal W}}$ and $\td{{\cal V}}$ respectively. They satisfy
\bea
&& 
0 \longrightarrow {\cal L}_1 \longrightarrow \td{{\cal W}}  \longrightarrow {\cal L}_2 \longrightarrow 0\,, 
\nonumber \\
&&
0 \longrightarrow \td{{\cal W}} \longrightarrow \td{{\cal V}}  \longrightarrow {\cal L}_3 \longrightarrow 0\,. 
\label{3.13}
\eea
Let us denote by $[\td{{\cal W}} ] $  and $[\td{{\cal V}}_{\td{{\cal W}}} ]$ the extension classes whose elements form the vector spaces
\be 
H^1 ({\cal A},  {\cal L}_1 \otimes {\cal L}_2^*)  = H^1({\cal A}, {\cal O}_{{\cal A}} (-2, 1, 2)) 
\qquad {\rm and} \qquad  H^1 ({\cal A},  \td{{\cal W}} \otimes {\cal L}_3^*)  \,. 
\label{3.14}
\ee
Similarly to our discussion in the previous subsection, we can introduce the moduli spaces of the corresponding vector bundles
\bea
&&
{\cal M}({\td{{\cal W}}}) = {\mathbb P}H^1 ({\cal A},  {\cal L}_1 \otimes {\cal L}_2^*) \,, \qquad \ \  {\rm dim} {\cal M}({\td{{\cal W}}} )=
h^1 ({\cal A},  {\cal L}_1 \otimes {\cal L}_2^*) -1\,, 
\nonumber \\
&&
{\cal M}({\td{{\cal V}}_{\td{{\cal W}}} })=  {\mathbb P}H^1 ({\cal A},  \td{{\cal W}} \otimes {\cal L}_3^*)\,, \qquad 
 {\rm dim} {\cal M}({\td{{\cal V}}_{ \td{{\cal W}}}})  =h^1 ({\cal A},  \td{{\cal W}} \otimes {\cal L}_3^*)-1 \,. 
\label{3.14.1}
\eea

Let us now study how the spaces~\eqref{3.14}, \eqref{3.14.1} 
are related to the similar spaces in~\eqref{3.6}, \eqref{3.6.2}, \eqref{3.7}.
The dimension of the first cohomology group in~\eqref{3.14} can be computed using the Kunneth formula and the Bott's formula
\be
\label{bott}
h^q(\mathbb{P}^n, {\cal O}_{{\mathbb P}^n}(k))=\begin{cases} \dfrac{(n+k)!}{n!k!} & {\rm for } \ q=0, \textrm{ } n \geq0 , \textrm{ } k\geq 0 \\  
\dfrac{(-k-1)!}{n! (-k-n-1)!} \textrm{ } & {\rm for } \ q=n, \textrm{ } n \geq 0,\textrm{ } k\leq -(n+1)\\ 0 & {\rm otherwise }  \end{cases}\,. 
\ee
We then find that 
\be
h^1 ({\cal A},  {\cal L}_1 \otimes {\cal L}_2^*) = 18\,. 
\label{3.15}
\ee
Comparing with~\eqref{3.8}, we observe that $ h^1 (\td{X}, L_1 \otimes L_2^*) = h^1 ({\cal A},  {\cal L}_1 \otimes {\cal L}_2^*) =18$. This means that 
all extensions in $ [\td{{\cal W}}]$ descend to non-trivial extensions  on 
$\td{X}$, and that all extensions in $[\td{ W}]$ are obtained as a restriction of extensions in $[ \td{{\cal W}}]$.
The dimension of the second cohomology group in~\eqref{3.14} can be obtained by tensoring the first line in~\eqref{3.13} with ${\cal L}_3^*$
to get
\be
0 \longrightarrow 
{\cal O}_{{\cal A}} (-4, 5, 1) \longrightarrow \td{{\cal W}}\otimes  {\cal L}_3^* \longrightarrow {\cal O}_{{\cal A}} (-2, 4, -1)\longrightarrow 0\,. 
\label{3.16}
\ee
Now the dimension of $H^1 ({\cal A},  \td{{\cal W}} \otimes {\cal L}_3^*)$ can be computed from the cohomology long exact sequence 
corresponding to~\eqref{3.16}, again using the Kunneth and Bott's formulas. We find that
\be 
h^1 ({\cal A},  \td{{\cal W}} \otimes {\cal L}_3^*) =189\,. 
\label{3.17}
\ee
Comparing this with eq.~\eqref{3.8}, we see that 
$h^1 ({\cal A},  \td{{\cal W}} \otimes {\cal L}_3^*)  > h^1 (\td{X},  \td{W} \otimes  L_3^*) $. 
This means that $189- 117 =72$  non-trivial extensions of  $[\td{{\cal V}}_{\td{{\cal W}}} ]$  on ${\cal A}$ get restricted 
to zero on $\td{X}$. 

Let us now describe the  space of extensions on $\td{X}$ in terms of the cohomology groups on the ambient space.
Cohomology groups on projective spaces can be written in terms of polynomials. Hence, this way we will obtain an explicit 
polynomial representation of the elements of  $[\td{W}]$  and $[\td{V}_{\td{W}}]$.
Furthermore, taking those polynomials which are invariant under ${\mathbb  Z}_3 \times {\mathbb  Z}_3$ will 
give us an explicit polynomial parametrization of $[W]$ and  $[V_W]$. Taking the projectivization of the corresponding vector spaces will 
give us a parametrization  of the moduli spaces  ${\cal M}({W})$ and ${\cal M}(V_W)$.

The relation between cohomology groups on ${\cal A}$  and on $\td{X} \subset { \cal A}$ can be obtained using the Koszul sequence--as we now explain. 
The Calabi-Yau threefold $\td{X}$ is defined as a submanifold in ${\cal A}$ using eqs.~\eqref{2.3}.
Since $\td{X}$ is of co-dimension 2, its normal bundle is a rank 2 vector bundle. From eqs.~\eqref{2.3} we find that it is a restriction 
of the following vector bundle on ${\cal A}$ :
\be 
{\cal N} = {\cal N}_1 \oplus  {\cal N}_2\,, \qquad  {\cal N}_1= {\cal O}_{{\cal A}} (1, 3, 0)\,, \quad  {\cal N}_2= {\cal O}_{{\cal A}} (1, 0, 3)\,. 
\label{3.18}
\ee
Let ${\cal L}$ be a vector bundle on ${\cal A}$ and $L= {\cal L}|_{\td{X}}$. They are related to each other by the Koszul sequence
\be 
0 \longrightarrow \wedge^2 {\cal N}^* \otimes {\cal L}  \stackrel{F'}{\longrightarrow} {\cal N}^*  \otimes {\cal L}  \stackrel{F}{\longrightarrow}  {\cal L}
\stackrel{r}{\longrightarrow} L  \longrightarrow 0\,. 
\label{3.19}
\ee
The map $r$ is the restriction map, the map $F$ is multiplication (from the left) by the row vector $ (F_1, F_2)$ of the defining polynomials in~\eqref{2.3}
and the map $F'$ is determined by the composition rule $ F \circ F'=0$, which follows from the exactness of~\eqref{3.19}. This implies that 
$F'$ is a column vector $(F_2, -F_1)^T$. 
Note that the sequence~\eqref{3.19} is not short and, hence, we cannot write the  long exact cohomology sequence directly. However, one can split~\eqref{3.19}
into two short exact sequences by introducing an auxiliary sheaf ${\cal S}$: 
\bea
&& 
 0 \longrightarrow \wedge^2 {\cal N}^* \otimes {\cal L}  \stackrel{F'}{\longrightarrow} {\cal N}^*  \otimes {\cal L} \stackrel{H_1}{\longrightarrow} {\cal S } \longrightarrow 0\,, 
 \nonumber \\
 &&
0 \longrightarrow  {\cal S }
\stackrel{H_2}{\longrightarrow}  {\cal L}
\stackrel{r}{\longrightarrow} L  \longrightarrow 0\,, 
\label{3.20}
\eea
where the maps $H_1, H_2$ satisfy $H_2 \circ H_1 =F$. Writing the long exact  cohomology sequences for~\eqref{3.20} 
allows us to compute the cohomology groups of $L$ in terms of the cohomology groups of ${\cal L}, {\cal N}^*  \otimes {\cal L}$ and $\wedge^2 {\cal N}^* \otimes {\cal L}$.
These, in turn, can be calculated using the Kunneth and Bott formulas. We will not present the details of these laborious calculations 
and only give the results. 

To compute the space of extensions  $[\td{W}]$ we apply the Koszul sequence~\eqref{3.18} to $L= L_1 \otimes L_2^*$. Then we find that 
\be 
H^1 (\td{X}, L_1 \otimes L_2^*) = 
H^1 ({\cal A}, {\cal L}_1 \otimes {\cal L}_2^*)= H^1 ({\cal A}, {\cal O}_{{\cal A}} (-2, 1, 2))\,. 
\label{3.21}
\ee
To compute the space of extensions  $[\td{V}_{\td{W}}]$, we apply the Koszul sequence~\eqref{3.19} to $L= \td{W} \otimes L_3^*$. 
But first we have to find the cohomology groups of $\td{{\cal W}} \otimes {\cal L}_3^*, \ 
\td{{\cal W}} \otimes {\cal L}_3^*\otimes {\cal N}^*$ and $\td{{\cal W}} \otimes {\cal L}_3^*\otimes \wedge^2 {\cal N}^*$ using~\eqref{3.13}.
We obtain
\be 
h^{\bullet} ({\cal A}, \td{{\cal W}} \otimes {\cal L}_3^*) = h^{\bullet}  ({\cal A}, {\cal O}_{{\cal A}} (-4, 5, 1))= (0, 189, 0, 0, 0, 0)\,, 
\label{3.22.1}
\ee
\bea
&&
h^{\bullet} ({\cal A}, \td{{\cal W}} \otimes {\cal N}^* \otimes {\cal L}_3^*) = (0, 72, 0, 36, 0, 0)\,, 
\nonumber \\
&&
h^1 ({\cal A}, \td{{\cal W}} \otimes {\cal N}^* \otimes {\cal L}_3^*)= h^{1}  ({\cal A}, {\cal O}_{{\cal A}} (-5, 2, 1))=72\,, \nonumber \\
&&
h^3 ({\cal A}, \td{{\cal W}} \otimes {\cal N}^* \otimes {\cal L}_3^*)= h^{3}  ({\cal A}, {\cal O}_{{\cal A}} (-3, 4, -4))=36\,,
\label{3.22.2}
\eea
\be
h^{\bullet} ({\cal A}, \td{{\cal W}} \otimes \wedge^2 {\cal N}^* \otimes {\cal L}_3^*)= 
h^{\bullet}  ({\cal A}, {\cal O}_{{\cal A}} (-4, 1, -4))= (0, 0, 0, 27, 0, 0)\,. 
\label{3.22.3}
\ee
Now, from the Koszul sequence~\eqref{3.19} applied to $L= \td{W} \otimes L_3^*$  we find 
\bea
&&
H^1 (\td{X},  \td{W} \otimes L_3^*) = \dfrac{H^1 ({\cal A}, \td{{\cal W}} \otimes {\cal L}_3^*)}{F_1 \cdot
H^1 ({\cal A}, \td{{\cal W}} \otimes {\cal N}^* \otimes {\cal L}_3^*)}  =
\dfrac{H^1 ({\cal A}, {\cal O}_{{\cal A}} (-4, 5, 1))}{F_1 \cdot H^1 ({\cal A}, {\cal O}_{{\cal A}} (-5, 2, 1))}\,, 
\nonumber \\
&& 
 h^1 (\td{X},  \td{W} \otimes L_3^*) = 189-72= 117\,. 
\label{3.23}
\eea
Here $F_1$ is  the first defining polynomial in~\eqref{2.3}. It can be viewed as an element of $H^0 ({\cal A}, {\cal O}_{{\cal A}} (1, 3, 0))$.
When we multiply $F_1$ by a differential in $H^1 ({\cal A}, {\cal O}_{{\cal A}} (-5, 2, 1))$, we naturally obtain a differential in 
$H^1({\cal A},  {\cal O}_{{\cal A}} (-4, 5, 1))$. Eq.~\eqref{3.23} simply means that to find the extension class $[\td{V}_{\td{W}}]$ on $\td{X}$ 
we have to mod out by the image of the map $F_1$. The elements in the image of $F_1$, that is, the denominator in~\eqref{3.23},
are precisely the extensions on the ambient space which do not correspond to extensions on $\td{X}$. 
From eq.~\eqref{3.23} we see that they become zero when we restrict to $\td{X}$, since $F_1$ vanishes on $\td{X}$. 

Note that the right hand side of eq.~\eqref{3.23} does not depend on the choice  of an element $\td{W}$ in $[\td{W}]$ or in
$ {\cal M}({\td{W}})$. 
This means that the moduli space ${\cal M}({\td{V}})$ is the trivial fibration
\be 
{\cal M}({\td{V}}) = {\cal M}({\td{W}} )\times {\cal M}({\td{V}_{\td{W}}})\,,  
\label{3.24}
\ee
where the first factor is the projectivization of the vector space in~\eqref{3.21} and the second factor is the projectivization 
of the vector space in~\eqref{3.23}. 


\subsection{Parametrization of the moduli space}
\label{moduli}


The aim of this subsection is to derive a parameterization of the moduli spaces ${\cal M}(\td{V}_{\td{W}})$ and 
${\cal M}(V_W)$. Parametrization of  the moduli spaces of extensions ${\cal M}(\td{W})$ and ${\cal M}(W)$
can be derived in a similar way, but is not required in this paper.

First, we consider the numerator in~\eqref{3.23}. According to the Kunneth and Bott formulas, 
\be
H^1 ({\cal A}, {\cal O}_{{\cal A}} (-4, 5, 1)) = H^1({\mathbb P}^1, {\cal O}_{{\mathbb P}^1} (-4) )\otimes
H^0({\mathbb P}^2 \times {\mathbb P}^2 , {\cal O}_{{\mathbb P}^2 \times {\mathbb P}^2} (5, 1)) \,. 
\label{3.25}
\ee
Let us consider the vector space 
\be 
H^1({\mathbb P}^1, {\cal O}_{{\mathbb P}^1} (-3) ) \simeq H^0({\mathbb P}^1, {\cal O}_{{\mathbb P}^1} (1) )^{*}\,,
\label{3.26}
\ee
where  we have used Serre duality. This vector space is 2-dimensional and we denote its basis as $\{r_0, r_1\}$.
This  basis is chosen to be dual to the basis  $\{t_0, t_1\}$ of homogeneous degree 1 polynomials on ${\mathbb P}^1$. The vector space of interest,
\be 
H^1({\mathbb P}^1, {\cal O}_{{\mathbb P}^1} (-4) )  \simeq H^0 ({\mathbb P}^1, {\cal O}_{{\mathbb P}^1} (2) )^* \ ,
\label{3.27}
\ee
is 3-dimensional  with a natural basis $\{r_0^2, r_0 r_1, r_1^2\}$
dual to the  basis $\{t_0^2, t_0 t_1, t_1^2\}$
of degree 2 polynomials on ${\mathbb P}^1$. It follows that an arbitrary element $v \in H^1 ({\cal A}, {\cal O}_{{\cal A}} (-4, 5, 1)) $ can be written as
\be 
v = r_0^2 f_1 ({\bf x}, {\bf y}) +r_0 r_1 f_2 ({\bf x}, {\bf y}) + r_1^2 f_3 ({\bf x}, {\bf y}) \,,
\label{3.28}
\ee
where $f_1, f_2, f_3$ are homogeneous polynomials on ${\mathbb P}^2 \times {\mathbb P}^2$ of degree $(5, 1)$. 
Here, to simplify our notation, we let ${\bf x}$ denote the coordinates on the first ${\mathbb P}^2$, ${\bf x} \equiv [x_0: x_1: x_2]$ and, similarly, 
${\bf y}$ denotes the coordinates on the second ${\mathbb P}^2$, ${\bf y} \equiv [y_0: y_1: y_2]$. The coefficients in the polynomials $f_1, f_2, f_3$ can be viewed as coordinates on 
$H^1 ({\cal A}, {\cal O}_{{\cal A}} (-4, 5, 1))$. As we computed in~\eqref{3.22.1}, there are 189 such coefficients. 

Since eventually we are interested 
in the moduli space of the vector bundle $V$ on $X$, we restrict $H^1 ({\cal A}, {\cal O}_{{\cal A}} (-4, 5, 1))$ to its  
subspace consisting of elements $v_{inv}$ which are invariant under ${\mathbb Z}_3 \times {\mathbb Z}_3$. 
Since the discrete group does not act on ${\mathbb P}^1$ (see eq.~\eqref{2.4}), the elements $r_0$ and $r_1$ are automatically invariant.
Hence, $v_{inv}$ is of the form~\eqref{3.28} where  the polynomials $f_1, f_2, f_3$ are restricted to be the ${\mathbb Z}_3 \times {\mathbb Z}_3$ invariant 
polynomials of degree $(5, 1)$.
Let us introduce a basis for these invariant polynomials:
%
\bea
&&
E_1= x_0^5 y_0 +x_1^5 y_1 +x_2^5 y_2\,, 
\nonumber \\
&& 
E_2= x_0^2 x_1^3 y_0 +x_1^2 x_2^3 y_1 +x_2^2 x_0^3 y_2\,, 
\nonumber \\
&& 
E_3= x_0^2 x_2^3 y_0 +x_1^2 x_0^3 y_1 +x_2^2 x_1^3 y_2\,, 
\nonumber \\
&& 
E_4= x_0^2 x_1 x_2 y_0 +x_1^2 x_2 x_0 y_1 +x_2^2 x_0 x_1 y_2\,, 
\nonumber \\
&& 
E_5= x_1^4 x_2 y_0 +x_2^4 x_0 y_1 +x_0^4 x_1 y_2\,, 
\nonumber \\
&& 
E_6= x_0^5 y_0 +x_1^5 y_1 +x_2^5 y_2\,, 
\nonumber \\
&& 
E_7= x_1 x_2^4 y_0 +x_2 x_0^4  y_1 +x_0 x_1^4 y_2\,.
\label{3.29}
\eea
The invariant polynomials $f_1, f_2, f_3$ are then given by 
\be 
f_1= \sum_{\a=1}^7 a_{\a} E_{\a}\,, \qquad 
f_2= \sum_{\a=1}^7 b_{\a} E_{\a}\,, \qquad f_3= \sum_{\a=1}^7 c_{\a} E_{\a}\,,
\label{3.30}
\ee
where $(a_{\a}, b_{\a}, c_{\a})$ are  coordinates on the 21(=189/9)-dimensional invariant subspace of $H^1 ({\cal A}, {\cal O}_{{\cal A}} (-4, 5, 1))$. 
However, to obtain the invariant part of the  extension class $[\td{V}_{ \td{W}}]$ we have to mod out by those 
elements which can be obtained by multiplying the defining polynomial $F_1$ by elements of $H^1 ({\cal A}, {\cal O}_{{\cal A}} (-5, 2, 1))$. 
We discuss this in detail in Appendix~\ref{D}. Here we simply state the result. Dividing by the the denominator in~\eqref{3.23} 
is equivalent to imposing the following constraints on the coordinates $(a_{\a}, b_{\a}, c_{\a})$:
\bea
&& 
a_1 +a_2 +a_3 =0 \,, \quad a_4+ a_5+ a_6=0 \,, 
\nonumber \\
&&
a_4 +b_1 +b_2 + b_3=0 \,, \quad a_7+ b_4+ b_5+ b_6=0 \,, 
\nonumber \\
&&
b_4 +c_1 +c_2 + c_3=0 \,, \quad b_7+ c_4+ c_5+ c_6=0 \,, 
\nonumber \\
&&
c_4=0\,, \qquad c_7=0\,. 
\label{3.31}
\eea
We can choose 
\be 
a_1, a_2, a_5, b_1, b_2, b_3, b_5, b_6, c_1, c_2, c_3, c_5, c_6
\label{3.32}
\ee
as independent parameters, with the others being determined using eqs.~\eqref{3.31}. 
The parameters in~\eqref{3.32} can be viewed 
as  the 13 coordinates on the invariant subspace of the vector space of extensions. Note that this is consistent with~\eqref{3.11}. According to our previous discussion, 
the moduli space  ${\cal M}({V_ W})$ is obtained by projectivization of this vector space. This simply means that we should view 
the coordinates  $(a_{\a}, b_{\a}, c_{\a})$ as homogeneous ones. 

To conclude this section, let us summarize the structure of the moduli space of $V$. It is given by the trivial fibration 
\be
{\cal M} (V)= {\cal M}(W) \times  {\cal M}(V_W)\,, 
\label{ins1}
\ee
where 
\be 
 {\cal M}(W) = {\mathbb P}^1\,, \qquad  {\cal M}(V_W)=  {\mathbb P}^{12}\,. 
 \label{ins2}
\ee
In total, we have 13 moduli of $V$. The parametrization of the second factor, ${\cal M}(V_W)$, is explicitly given by 
21 homogenous coordinates $(a_1, \dots, a_7, b_1, \dots, b_7, c_1, \dots, c_7)$ subject to 8 linear constraints~\eqref{3.31}. 
We can choose 13 independent variables, as in eq.~\eqref{3.32}, and view them  as 13 homogeneous coordinates on ${\cal M}(V_W)$. 
By similar methods, we can obtain a parametrization of  ${\cal M}(W)$--but we do not need it in this paper. 


\section{The superpotential on $X$}
\label{calcul}


In this section, we will compute the leading non-perturbative superpotential, that is, $\sim p= e^{i T^1}$ (see eq.~\eqref{2.17.2}), in a heterotic string vacuum specified  by $(X, V)$. 
To simplify our analysis, we will perform the calculations for fixed complex structure. Then $({\rm det} {\bar \pt}_{{\cal O}_{C_i} (-1)})$ in~\eqref{1.11}
become numerical constants which will not play any role and will be ignored. Our aim in this section will be to compute the Pfaffians. 
First, we will calculate them on the covering space $\td{X}$ and then on the quotient space $X$. Since we would like to compare these 
two calculations, in the theory on $\td{X}$  we will restrict ourselves to the invariant component of the moduli space which will descend to $X$.
The method of computing the Pfaffians will be similar to the one introduced in~\cite{Buchbinder:2002ic, Buchbinder:2002pr}. 
Since we do not know either the metric or the connection, we will rely on an algebraic approach whose essence is 
to understand under which conditions each Pfaffian vanishes. The conditions will be derived as a homogeneous polynomial equation on the moduli space.
Since the Pfaffian is a section of a line bundle on the moduli space and the moduli space is a projective space, this polynomial will be the Pfaffian up to a 
numerical coefficient which cannot be determined by our algebraic method. 

Let us now review the general condition for the vanishing of a Pfaffian on a holomorphic, isolated, genus 0 curve $C$~\cite{Witten:1996bn, Buchbinder:2002ic, Buchbinder:2002pr}. 
The Pfaffian vanishes if and only if the operator ${\bar \pt}_{V_{C} (-1)}$ has a zero mode. The zero modes of $\bar \pt$
are elements of the Dolbeault cohomology group. In the present case, the cohomology group of interest is $H^0 (C, V|_{C} \otimes {\cal O}_C (-1))$. 
Hence, the Pfaffian vanishes if and only if $h^0 (C, V|_{C} \otimes {\cal O}_C (-1)) \neq 0$. 
Since $h^0 (C, V|_{C} \otimes {\cal O}_C (-1)) $ is not a topological invariant, it depends on where we are in the moduli space of $V$. 
For generic values of the moduli, $h^0 (C, V|_{C} \otimes {\cal O}_C (-1)) $ will be zero and ${\bar \pt}_{V_{C} (-1)}$ will not have zero modes. 
However, at a specific co-dimension 1 subspace of the moduli space $h^0 (C, V|_{C} \otimes {\cal O}_C (-1)) $  will jump--thus producing a zero mode.
The Pfaffian of ${\bar \pt}_{V_{C} (-1)}$ will be determined by the equation defining this co-dimension 1 subspace. 


\subsection{Calculation of the Pfaffians}
\label{calcultx}


As was discussed in Section~\ref{curves}, on $\td{X} $  there are 81 isolated curves of interest. These curves split into 9 orbits under the action of 
${\mathbb Z}_3 \times {\mathbb Z}_3$  with 9 curves in each orbit. If we restrict ourselves to the invariant part of the moduli space,
all curves in the same orbit will give an identical contribution. Hence, in this case we need to compute the Pfaffians of the Dirac operator on
any 9 curves which do not lie in the orbits of each other. An example of such curves was given in eqs.~\eqref{2.27}, \eqref{2.27.1}, \eqref{2.8}. 
Let us recall that these curves lie in different homology classes and, hence, have different areas measured by the Kahler form on $\td{X}$. 
However, they have the same area when measured using the invariant part of the Kahler form $\o_{{\cal A}}|_{\td{X}}$. Therefore,
the images of these curves in $X$  have the same area
with respect to the Kahler form  $\o_X$. 

Let us now study under which conditions $h^0 (C, \td{V}|_{C} \otimes {\cal O}_C (-1)) \neq 0$ for the curves $C$ of the type 
${\mathbb P}^1 \times {\bf x} \times {\bf y} \subset \td{X} \subset  {\mathbb P}^1 \times {\mathbb P}^2 \times {\mathbb P}^2$.
We denote ${\cal B}= {\mathbb P}^2 \times {\mathbb P}^2$ and define $p_{{\cal B}}$ to be the projection 
$p_{{\cal B}}: {\cal A} \to  {\cal B}$ with fibers being ${\mathbb P}^1$. Now consider a particular extension element $\td{{\cal  W}}$ 
on ${\cal A}$, 
\be 
0 \longrightarrow {\cal L}_1 \longrightarrow \td{{\cal W}}  \longrightarrow {\cal L}_2 \longrightarrow 0 \ ,
\label{4.2}
\ee
and tensor this sequence with ${\cal O}_{{\cal A}}(-1, 0, 0)$ to obtain
\be 
0 \longrightarrow {\cal O}_{{\cal A}}( -3, 2, 1) \longrightarrow \td{{\cal W}}(-1, 0, 0)  \longrightarrow  {\cal O}_{{\cal A}}( -1, 1, 1) \longrightarrow 0\,,
\label{4.3}
\ee
where we have used eqs.~\eqref{3.12} and defined $\td{{\cal W}}(-1, 0, 0) =\td{{\cal W}} \otimes {\cal O}_{{\cal A}}( -1, 0, 0)$. Now take  the direct image 
of this sequence with the projection $p_{\cal B}$. This leads to the exact sequence
\bea
 0 \longrightarrow &p_{{\cal B }*} {\cal O}_{{\cal A}}( -3, 2, 1) & 
 \longrightarrow p_{{\cal B }*} \td{{\cal W}}(-1, 0, 0)  \longrightarrow p_{{\cal B }*}  {\cal O}_{{\cal A}}( -1, 1, 1) \longrightarrow 
 \nonumber \\
&R^1 p_{{\cal B }*} {\cal O}_{{\cal A}}( -3, 2, 1) & 
 \longrightarrow R^1 p_{{\cal B}*} \td{{\cal W}}(-1, 0, 0)  \longrightarrow R^1 p_{{\cal B}*}  {\cal O}_{{\cal A}}( -1, 1, 1) \longrightarrow 0\,. 
 \label{4.4}
\eea
At each point on ${\cal B}$ and  for any line bundle ${\cal L}$,  $p_{{\cal B }*}  {\cal L}$ is generated by the cohomology group of the fiber at this point; that is by
$H^0 ({\mathbb P}^1, {\cal L}|_{{\mathbb P}^1})$. 
Similarly,  $R^1 p_{{\cal B }*}  {\cal L}$ is generated by 
$H^1 ({\mathbb P}^1, {\cal L}|_{{\mathbb P}^1})$. Clearly, for ${\cal L}$ of the form ${\cal L}= {\cal O}_{{\cal A}}(m_1, m_2, m_3)$, we have 
${\cal L}|_{{\mathbb P}^1}  = {\cal O}_{{\mathbb P}^1}(m_1)$. Then, 
using the Bott's formula, we compute that 
\bea 
&&
p_{{\cal B }*} {\cal O}_{{\cal A}}( -3, 2, 1)= 0\,, \quad p_{{\cal B }*}  {\cal O}_{{\cal A}}( -1, 1, 1)=0\,, \quad R^1 p_{{\cal B}*}  {\cal O}_{{\cal A}}( -1, 1, 1)=0\,,  
\nonumber \\
&&
R^1 p_{{\cal B }*} {\cal O}_{{\cal A}}( -3, 2, 1)= H^1({\mathbb P}^1,  {\cal O}_{{\mathbb P}^1}(-3)) \otimes {\cal O}_{{\cal B}}(2, 1 )\,. 
\label{4.5}
\eea
Therefore, from eq.~\eqref{4.4} we obtain
\bea
&&
p_{{\cal B }*} \td{{\cal W}}(-1, 0, 0) =0\,, \nonumber \\
&&
 R^1 p_{{\cal B}*} \td{{\cal W}}(-1, 0, 0) = 
R^1 p_{{\cal B }*} {\cal O}_{{\cal A}}( -3, 2, 1)= H^1({\mathbb P}^1,  {\cal O}_{{\mathbb P}^1}(-3)) \otimes {\cal O}_{{\cal B}}(2, 1 )\,. 
\label{4.6}
\eea
Note that the right hand side in eqs.~\eqref{4.6} is independent of the choice of the extension representative $\td{\cal W}$.

As the next step, we consider the sequence defining $\td{\cal V}$ tensored with $ {\cal O}_{{\cal A}}( -1, 0, 0)$. Using eqs.~\eqref{3.12}, \eqref{3.13} 
we obtain
\be 
0 \longrightarrow \td{{\cal W}}(-1, 0, 0) \longrightarrow \td{{\cal V}}(-1, 0, 0)  \longrightarrow  {\cal O}_{{\cal A}}( 1, -3, 0) \longrightarrow 0\,,
\label{4.7}
\ee
where  $\td{{\cal V}}(-1, 0, 0) =\td{{\cal V}} \otimes {\cal O}_{{\cal A}}( -1, 0, 0)$. Taking the direct image of this sequence with the projection 
$p_{{\cal B}}$ and using~\eqref{4.6} we obtain 
\bea
0 \longrightarrow & p_{{\cal B}*} \td{{\cal V}}(-1, 0, 0)  &   \longrightarrow 
\nonumber \\
&H^0({\mathbb P}^1,  {\cal O}_{{\mathbb P}^1}(1)) \otimes {\cal O}_{{\cal B}}(-3, 0 )&  \stackrel{\d(\td{{\cal V}})}{\longrightarrow }
H^1({\mathbb P}^1,  {\cal O}_{{\mathbb P}^1}(-3)) \otimes {\cal O}_{{\cal B}}(2, 1 ) \longrightarrow 
\nonumber \\
& R^1 p_{{\cal B}*} \td{{\cal V}}(-1, 0, 0)  &  \longrightarrow  0\,. 
\label{4.8}
\eea
Here $\d(\td{{\cal V}})$ is a map depending on the moduli of $\td{{\cal V}}$. Note that the vector spaces $H^0({\mathbb P}^1,  {\cal O}_{{\mathbb P}^1}(1))$ 
and $H^1({\mathbb P}^1,  {\cal O}_{{\mathbb P}^1}(-3)) $ are 2-dimensional, which implies  that $\d(\td{{\cal V}})$ can be represented by a $2 \times 2$ matrix. 

Let us now consider  the sequence~\eqref{4.8} at any point ${\bf x} \times {\bf y}  \subset {\cal B}$ which corresponds to a curve
$C = {\mathbb P}^1 \times {\bf x} \times {\bf y} \subset  {\cal A}$. This curve is also a curve in $\td{X}$ if the point ${\bf x} \times {\bf y}$
satisfies equations~\eqref{2.23}.
At any point $ {\bf x} \times {\bf y} \in {\cal B} $,  $p_{{\cal B}*} \td{{\cal V}}(-1, 0, 0)$ 
is generated by the cohomology group $H^0 (C, \td{{\cal V}}|_C \otimes {\cal O}_C (-1))$. 
If ${\bf x} \times {\bf y}$ is chosen to satisfy eq.~\eqref{2.23}, then we also have 
$H^0 (C, \td{{\cal V}}|_C \otimes {\cal O}_C (-1))=H^0 (C, \td{V}|_C \otimes {\cal O}_C (-1))$--which is precisely the space of  zero modes of the Dirac operator. 
This space is non-empty if and only if the map $\d(\td{{\cal V}})|_{{\bf x} \times {\bf y}}$ has a non-trivial kernel. 
Let us define $\d(\td{V}) = \d(\td{{\cal V}})|_{\td{X}}$.
Note that if ${\bf x} \times {\bf y}$ corresponds to a  curve in $\td{X}$, then we have $\d(\td{V})|_{{\bf x} \times {\bf y}} =\d(\td{{\cal V}})|_{{\bf x} \times {\bf y}}$.
If we represent $\d(\td{V})|_{{\bf x} \times {\bf y}}$  by a $2 \times 2 $ matrix, it then follows that 
\be 
{\rm Pfaff}_{\td{X}}  ({\bar \pt}_{V_{C_i} (-1)}) =0  \qquad {\rm if \ and \ only \ if} \qquad det [\d(\td{V})|_{{\bf x} \times {\bf y}}] =0\,. 
\label{4.9}
\ee
Let us now construct the matrix $\d(\td{V})$ which provides a map $\d(\td{V}): {\cal H}_1 \to {\cal H}_2$, where we denoted
\bea
&& 
{\cal H}_1=  H^0({\mathbb P}^1,  {\cal O}_{{\mathbb P}^1}(1)) \otimes {\cal O}_{{\cal B}}(-3, 0 ) \,, 
\nonumber \\
&&
{\cal H}_2= H^1({\mathbb P}^1,  {\cal O}_{{\mathbb P}^1}(-3)) \otimes {\cal O}_{{\cal B}}(2, 1 )\,. 
\label{4.10}
\eea
Recall from eq.~\eqref{3.23} that modulo the denominator--which will be taken into account later--the space of extensions 
is given by the elements 
\be
v \in  H^1 ({\cal A}, {\cal O}_{{\cal A}} (-4, 5, 1)) = H^0({\mathbb P}^1,  {\cal O}_{{\mathbb P}^1}(-4)) \otimes 
H^1({\cal B}, {\cal O}_{{\cal B}}(5,1 ))\,, 
\label{4.11}
\ee
where $v$ can explicitly be written as (see~\eqref{3.28})
\be 
v = r_0^2 f_1 ({\bf x}, {\bf y}) +r_0 r_1 f_2 ({\bf x}, {\bf y}) + r_1^2 f_3 ({\bf x}, {\bf y}) \,.
\label{4.12}
\ee
Comparing eq.~\eqref{4.10} and eq.~\eqref{4.11}, we conclude that $\d(\td{V})$ is given by multiplication by $v$. As was discussed around eq.~\eqref{3.26},
we can introduce the basis  $\{t_0, t_1\}$ for $H^0({\mathbb P}^1,  {\cal O}_{{\mathbb P}^1}(1)) $ and the dual basis $\{r_0, r_1\}$ for 
$H^1({\mathbb P}^1,  {\cal O}_{{\mathbb P}^1}(-3)) $. To construct the matrix $\d(\td{V})$, we simply multiply $v$ by the basis
elements $\{t_0, t_1\}$ to get 
\be  
v (t_0)= r_0 f_1 + r_1 f_2\,, \qquad  v (t_1)= r_0 f_2 + r_1 f_3
\label{4.13}
\ee
and present the answer in the  matrix form
\begin{eqnarray}
\d(\td{V})  \begin{pmatrix} r_0   \\ r_1   \end{pmatrix} \,. 
\label{4.14}
\end{eqnarray}
This gives 
\begin{eqnarray}
\d(\td{V}) =  \begin{pmatrix} f_1 & f_2   \\ f_2 & f_3   \end{pmatrix} \,, \qquad  \det [\d(\td{V})] = f_1 f_3 - f_2^2\,. 
\label{4.15}
\end{eqnarray}
If $C$ is a curve corresponding to a specific point ${\bf x}\times {\bf y} \in {\mathbb P}^2\times {\mathbb P}^2$ then we get
\be 
det [\d(\td{V})|_{{\bf x} \times {\bf y}}]  =  (f_1 f_3 - f_2^2) ({\bf x},  {\bf y})\,. 
\label{4.16}
\ee
After we evaluate $f_1 f_3 - f_2^2$  at points of $ {\mathbb P}^2\times {\mathbb P}^2$,  the right hand side of~\eqref{4.16}
becomes a degree 2 homogeneous polynomial of the parameters of $f_1, f_2, f_3$. 
For the purposes of our paper, we can restrict $f_1, f_2, f_3$  to be the invariant polynomials under ${\mathbb Z}_3 \times {\mathbb Z}_3$. 
They are explicitly given by eqs.~\eqref{3.29}, \eqref{3.30}. Furthermore, we recall that to describe the extensions of $\td{V}$ rather than those
of $\cal{\td{V}}$, we have to impose the relations~\eqref{3.31}. 
We also recall that to describe the moduli space of $\cal{\td{V}}$, we projectivize the corresponding space of extensions.
This simply means that we view the parameters $(a_{\a}, b_{\a}, c_{\a})$ of the polynomials  $f_1, f_2, f_3$ as homogeneous coordinates.
Since the moduli space is a projective space and $(f_1 f_3 - f_2^2) ({\bf x},  {\bf y})$ is a homogeneous polynomial of degree 2,
we conclude that~\eqref{4.16} is a section of a line bundle of degree 2 on the moduli space. 
Finally, we notice that eq.~\eqref{4.16} depends only on the coordinates of ${\cal M}(\td{V}_{\td{W}})$. The coordinates of 
 ${\cal M}(\td{W})$ drop out from our calculations. 
 
From eq.~\eqref{4.9} and the fact the Pfaffian is a section of a line bundle on the moduli space, we conclude that 
\be
{\rm Pfaff}_{\td{X}}  ({\bar \pt}_{V_{C} (-1)}) \sim  (f_1 f_3 - f_2^2) ({\bf x},  {\bf y}) =\sum_{\a, \b=1}^7 (a_{\a} c_{\b}- b_{\a} b_{\b})
E_{\a} E_{\b} ({\bf x},  {\bf y}) 
\label{4.17}
\ee
up to a numerical coefficient--which we are not able to compute by our method. 
Let us now apply this result to the curves~\eqref{2.27}, \eqref{2.28}. Denote
\be 
{\cal R}_{{\td X}, i} = (f_1 f_3 - f_2^2) (s_i)\,, \qquad i=1, \dots, 9\,. 
\label{4.18}
\ee
Substituting the points~\eqref{2.27}, \eqref{2.28} into~\eqref{4.18}, we obtain the following expressions for ${\cal R}_i$ ($\zeta= e^{2 \pi i/3}$):
\bea 
&&
{\cal R}_{{\td X}, 1}  =-(2b_1-b_2-b_3)^2 + (2a_1-a_2-a_3) (2c_1-c_2-c_3)\,,
\nonumber 
%
%
\\
&&
{\cal R}_{{\td X}, 2}  = -(b_2 + b_3 \zeta^2 + b_1 \zeta)^2 + (a_2 +a_3 \zeta^2 + a_1 \zeta) (c_2 +c_3 \zeta^2 + c_1 \zeta)\,,
\nonumber
\\
&& 
{\cal R}_{{\td X}, 3}  =-(b_2 +b_3 \zeta +b_1 \zeta^2)^2 + (a_2 +a_3 \zeta +a_1 \zeta^2) (c_2 +c_3 \zeta^2 + c_1 \zeta)\,,
\nonumber
\\
&&
{\cal R}_{{\td X}, 4}  =-(-b_1+b_3+b_5-b_6)^2 + (-a_1+a_3+a_5-a_6) ( -c_1+c_3+c_5-c_6)\,,
\nonumber
\\
&&
{\cal R}_{{\td X}, 5}  =-(-b_1+b_2-b_5+b_6)^2 + (-a_1+a_2-a_5+a_6) ( -c_1+c_2-c_5+c_6)\,,
\nonumber
\\
&&
{\cal R}_{{\td X}, 6} =- (-b_1+ b_3 + (b_5-b_6) \zeta^2)^2 + (-a_1+ a_3 + (a_5-a_6) \zeta^2) (-c_1+ c_3 + (c_5-c_6) \zeta^2)\,, 
\nonumber
\\
&&
{\cal R}_{{\td X}, 7}  =- (-b_1+ b_2 - (b_5-b_6) \zeta^2)^2 + (-a_1+ a_2 - (a_5-a_6) \zeta^2) (-c_1+ c_2 - (c_5-c_6) \zeta^2)\,,
\nonumber
\\
&&
{\cal R}_{{\td X}, 8}  =- (-b_1+ b_2 - (b_5-b_6) \zeta)^2 + (-a_1+ a_2 - (a_5-a_6) \zeta) (-c_1+ c_2 - (c_5-c_6) \zeta)\,,
\nonumber
\\
&& 
{\cal R}_{{\td X}, 9}  =- (-b_1+ b_3 + (b_5-b_6) \zeta)^2 +(-a_1+ a_3 + (a_5-a_6) \zeta) (-c_1+ c_3 + (c_5-c_6) \zeta)\,.
\label{r19}
\eea
The parameters $(a_{\a}, b_{\a}, c_{\a})$ satisfy the relations~\eqref{3.31}, but substituting them into~\eqref{r19} 
does not lead to a simplification. Note that none of the polynomials ${\cal R}_{{\td X}, i}$ depends on the parameters 
$a_4, a_7, b_4, b_7, c_4, c_7$. The reason is because the corresponding polynomials $E_4$ and $E_7$ in~\eqref{3.29}
vanish on any curve satisfying eqs.~\eqref{2.23}.

Let us now introduce  the proportionality coefficient into~\eqref{4.17}. For each of our 9 curves, we denote it by $A_{\td{X}, i}$ where $i=1,\dots,9$. 
That is, we have 
\be
{\rm Pfaff}_{\td{X}}  ({\bar \pt}_{\td{V}_{C_i} (-1)})  = A_{\td{X}, i} {\cal R}_{\td{X}, i}\,. 
\label{4.19}
\ee
Note that every $A_{\td{X}, i}$ is non-zero because the Pfaffians vanish only along the zero locus of the polynomials
${\cal R}_{\td{X}, i}$ and do not vanish identically. We are not able to compute the coefficients $A_{\td{X}, i}$ by our algebraic method. 
However, it is possible to constrain them using the Beasley-Witten residue theorem, which we will now discuss. 


\subsection{The residue theorem on $\td{X}$}

Our theory on $\td{X}$ formally satisfies
the assumptions of Beasley and Witten in~\cite{Beasley:2003fx}, which we reviewed in Subsection~\ref{BW}. 
Indeed, the Calabi-Yau threefold $\td{X}$ is, by construction, a projective complete intersection manifold and the vector bundle $\td{V}$ is 
the restriction of a vector bundle $\cal{\td{V}}$. 
Nevertheless, as we discussed in Subsection~\ref{assumption}, the residue theorem of Beasley-Witten is not directly applicable here
since $h^{1, 1} (\td{X}) > h^{1,1} ({\cal A})$. However, indirectly we can still apply it. If we measure the area of curves in $\td{X}$ 
using the $(1, 1)$ form $\o_{{\cal A}}|_{\td{X}}$ then, according to the residue theorem,  the sum of the Pfaffians of all curves with the same area has to vanish. 
The 81 curves found in Subsection~\ref{curves} have the same area with respect to $\o_{{\cal A}}|_{\td{X}}$ and, hence, we can apply the residue theorem to them.
Since we are restricting ourselves to the invariant part of the moduli space, all curves in the same ${\mathbb Z}_3 \times {\mathbb Z}_3$
orbit will have an identical Pfaffian. This means that it is enough to sum the Pfaffians of the curves which do not lie in the orbit of each other. 
These Pfaffians are given in eqs.~\eqref{r19} and~\eqref{4.19}. Hence, the residue theorem implies that
\be 
 \sum_{i=1}^{9}
 A_{\td{X}, i} {\cal R}_{\td{X}, i} =0\,. 
\label{4.20}
\ee
Let us  stress again that eq.~\eqref{4.20} does not imply that the superpotentail in the heterotic theory on $\td{X}$ vanishes
because in~\eqref{4.20} we are summing the Pfaffians of curves lying in different homology classes and having different area with respect to the proper 
Kahler form $\o_{\td{X}}$. Hence, in the superpotential these Pfaffians will be weighted with different exponential prefactors and cannot cancel each other.
Eq.~\eqref{4.20}  constrains the coefficients $A_{\td{X}, i} $. It is possible to satisfy eq.~\eqref{4.20} if and only if 
the polynomials ${\cal R}_{\td{X}, i}$ in~\eqref{r19} are linearly dependent--which is a  non-trivial 
consistency check on our calculations. It is possible to check (using e.g. MATHEMATICA) that these polynomials are indeed linearly dependent 
and it possible to adjust the parameters $A_{\td{X}, i}$ so that the sum in~\eqref{4.20} vanishes. More precisely, requiring that the sum in~\eqref{4.20} 
vanish puts the following constraints on $A_{\td{X}, i}$:
\bea
&&
 A_{\td{X}, 1}=- A_{\td{X}, 4} - A_{\td{X}, 5}\,, \nonumber \\
 &&
    A_{\td{X}, 2} =e^{i \pi/3}A_{\td{X}, 4}-A_{\td{X}, 7}\,, \nonumber \\
    &&
A_{\td{X}, 3} =-e^{i \pi/3}A_{\td{X}, 5}+e^{-i \pi/3} (A_{\td{X}, 4}-A_{\td{X}, 7})\,, 
\nonumber \\
&&
 A_{\td{X}, 6} = A_{\td{X}, 4} +A_{\td{X}, 5} -A_{\td{X}, 7}\,, \nonumber \\
&&  
  A_{\td{X}, 8} = e^{i \pi/3}A_{\td{X}, 5} -e^{2i \pi/3}A_{\td{X}, 7} \,, \nonumber \\
  &&
A_{\td{X}, 9} = A_{\td{X}, 4}+e^{-i \pi/3} (A_{\td{X}, 5}-A_{\td{X}, 7})\,. 
 \label{4.21}
\eea
These results will play a role in constructing the superpotential  on the quotient Calabi-Yau manifold $X$.


\subsection{The explicit formula for the superpotential on $X$}
\label{calculx}


Since the curves $C_i$ corresponding to solutions~\eqref{2.27}, \eqref{2.28} lie in  different ${\mathbb Z}_3 \times {\mathbb Z}_3$ orbits, 
they descend to 9 different curves on $X$. To simplify our notation, we will still denote these curves by $C_i$. The non-perturbative superpotential 
for these curves is then given by 
\be 
W_{X} ([C])=  e^{i T^1} \sum_{i=1}^{9}
{\rm Pfaff}_{X}  ({\bar \pt}_{V_{C_i} (-1)}) \chi_i \,. 
\label{4.22}
\ee
As was discussed in section~\ref{CY}, these 9 curves lie in the same homology class in $H_2 (X, {\mathbb R})$ but in 9 different 
homology classes in $H_2 (X, {\mathbb Z})$. This means that the contribution for each curve will pick up a distinct ${\mathbb Z}_3 \times {\mathbb Z}_3$
character $\chi_i$. As we discussed in Subsection~\ref{tor}, as long as the curves in the same homology class in  $H_2 (X, {\mathbb Z})$ 
receive the same character and curves in different homology classes receive  different characters,
the distribution of the characters among the curves is arbitrary
and depends on the choice of the connected component of the moduli space of the $B$-field.

To compute $W_{X} ([C])$ in~\eqref{4.22}, we notice that since the Pfaffians in the previous subsection  were computed for the invariant part of the moduli space.
It follows that
\be
{\rm Pfaff}_{X}  ({\bar \pt}_{V_{C_i} (-1)}) = {\rm Pfaff}_{\td{X}}  ({\bar \pt}_{\td{V}_{C_i} (-1)}) \,, 
\label{4.23}
\ee
where the right hand side is given by eqs.~\eqref{4.19}, \eqref{r19}, \eqref{4.21}. In particular, it follows that 
\be 
{\rm Pfaff}_{X}  ({\bar \pt}_{V_{C_i} (-1)}) =A_{X, i} {\cal R}_{X, i}\,, \qquad  {\cal R}_{{X, i}}= {\cal R}_{\td{X}, i}\,, \quad A_{X, i}=A_{\td{X}, i}\,. 
\label{4.24}
\ee
The reason is that, by construction, the gauge connection on $V$ is the equivariant connection on $\td{V}$. 
For the trivial choice of the equivariant structure, it is just the connection on $\td{V}$ restricted to the invariant part of the moduli space. 
Since  in the previous subsection we restricted our calculations to the invariant part of the moduli space, the Dirac operators on
both sides in~\eqref{4.23} depend on the same connection and, hence, their Pfaffians are equal. Thus, the superpotential in~\eqref{4.22}
becomes
\be 
W_{X} ([C])=  e^{i T^1} \sum_{i=1}^{9}  \chi_i A_{X, i} {\cal R}_{X, i}\,, 
\label{4.25}
\ee
where ${\cal R}_{X, i}$ are also given by~\eqref{r19}  and $A_{X, i} $ also satisfy the constraints~\eqref{4.21}. 
Now, the key observation is that, since the linear combination $\sum_{i=1}^{9} A_{X, i} {\cal R}_{X, i}=0$ due to the residue theorem,
the linear combination in~\eqref{4.25}
$\sum_{i=1}^{9}  \chi_i A_{X, i} {\cal R}_{X, i}$ is non-zero because it is twisted by the characters $\chi_i$, most of which are not unity. 

As an example let us give some assignment of the characters to the curves. Note that though 9 different characters $\chi_i$ label 9 inequivalent 
representations of  ${\mathbb Z}_3 \times {\mathbb Z}_3$, they can take only 3 values in ${\mathbb C}^*$ given by
$1, e^{2 \pi i/3}, e^{4 \pi i/3}$.  Assigning, for example, 
\be 
\chi_1=\chi_2=\chi_3=1\,  \quad \chi_4=\chi_5=\chi_6=e^{2 \pi i/3}\,, \quad  \chi_7=\chi_8=\chi_9=e^{4 \pi i/3}
\label{4.26}
\ee
we obtain 
\be 
W_{X} ([C])=  e^{i T^1} \big(\sum_{i=1}^{3} A_{X, i} {\cal R}_{X, i} 
+ e^{2 \pi i/3}  \sum_{i=4}^{6}  A_{X, i} {\cal R}_{X, i} +e^{4 \pi i/3}
\sum_{i=7}^{9}  {\cal R}_{X, i}  \big)\,.
\label{4.27}
\ee
It is straightforward to check using eqs.~\eqref{r19} and~\eqref{4.21} that the linear combination~\eqref{4.27} does not vanish. 
From eqs.~\eqref{3.31}, \eqref{3.32}, \eqref{r19} we see that $W_{X} ([C])$ depends on 12 out of the 13 moduli of ${\cal M}(V)$. 
The remaining modulus parametrizing ${\cal M}(W)$ does not show up in the superpotential~\eqref{4.27}. 
Our expression for $W_{X} ([C])$ depends on 9 numerical coefficients $A_{X, i}$ which we cannot fully compute by our algebraic method. 
However, due to relations~\eqref{4.21}, only 3 of them are really unknown.  

Thus, we have explicitly demonstrated that in our  model on $X$ a non-vanishing, non-perturbative superpotential can be generated
in the low-energy field theory. 


\section{Conclusion and future directions}


In this paper, we presented  examples of heterotic string compactifications  with non-vanishing non-perturbative superpotentials.
In our examples, the superpotential does not vanish on both the simply connected covering space and the non-simply connected
manifold obtained as  a quotient by the action of the discrete isometry group. In both cases, the reason for the non-vanishing of the superpotential 
can be attributed to the existence of holomorphic, isolated, genus 0 curves which are unique in their integer homology classes. 

It would be interesting to generalize the ideas developed in this paper for realistic heterotic models and to compute non-perturbative 
superpotentials in a heterotic MSSM. The heterotic Standard Model constructed in~\cite{Braun:2005ux, Braun:2005bw, Braun:2005zv, Braun:2005nv} 
used a different Schoen manifold with a different action of  ${\mathbb Z}_3 \times {\mathbb Z}_3$. Hence, it would be interesting to see if one can build a 
heterotic MSSM on the Schoen manifold used in this paper. Then one can extend the results of this paper to compute the non-perturbative 
superpotential in an MSSM, rather than in a toy model. The result  is expected to be non-zero, as in our present examples. 

Another possible direction is to apply our methods to realistic heterotic models obtained using the 
monad construction~\cite{Anderson:2007nc, Anderson:2009mh, Buchbinder:2013dna, Buchbinder:2014qda, Buchbinder:2014sya, Anderson:2008uw}. 
The crucial difference is that such models are built on projective Calabi-Yau manifolds satisfying $h^{1, 1} (\td{X}) = h^{1, 1} ({\cal A})$. 
Then, according to the Beasley-Witten residue theorem, the non-perturbative superpotential vanishes on the covering manifold $\td{X}$. 
However, on the quotient manifold $X$ it might be non-zero because the second homology group of $X$ is expected to contain discrete torsion. 
It would be interesting to see if one indeed can generate a non-perturbative superpotential in such models.


\section*{Acknowledgements}
The authors are very grateful to Tony Pantev for valuable discussions. The authors would also like to thank Ling Lin
for helpful conversations. The work of E. I. Buchbinder was supported by the ARC Future Fellowship FT120100466
and in part by the ARC Discovery project DP140103925.
B. A. Ovrut is supported in part by the DOE under contract No. DE-SC0007901.
E.I.B.~would like to thank the physics department at the University of Pennsylvania 
where some of this work was done for warm hospitality.
\newpage


\appendix
\section{The normal bundle to the curves in $\td{X}$}
\label{A}


Here we will compute the normal bundle to the curves in Subsection~\ref{curves}. Specifically, we present our calculations for the curve specified by $s_1$ in eq.~\eqref{2.27}. 
The other curves can be treated similarly and give the same result. 

The curve $s_1$ is of the form 
\be
C_1 =[t_0:t_1]\times s_1 = [t_0, t_1] \times [1 :-1:0] \times [1, -1, 0]\,. 
\label{A1}
\ee
Let us first consider the short exact sequence relating the tangent bundle $T\td{X}$ and the normal bundle $N \td{X}$  of $\td{X}$; that is
\be
0 \longrightarrow T\td{X} \stackrel{h_2}{\longrightarrow} T {\cal A}|_{\td{X}} \stackrel{h_1}{\longrightarrow} N \td{X} \longrightarrow 0\,, 
\label{A2}
\ee
where $T {\cal A}$ is the tangent bundle of the ambient space given by 
\be 
T {\cal A} = {\cal O}_{{\mathbb P}^1} (2) \oplus T{\mathbb P}^2 \oplus T{\mathbb P}^2
\label{A3}
\ee
and we have used the fact that the tangent bundle of ${\mathbb P}^1$ is ${\cal O}_{{\mathbb P}^1} (2)$.
Using eqs.~\eqref{2.3}, the normal bundle $N\td{X}$ is 
\be 
N\td{X} = {\cal O}_{{\cal A}} (1, 3, 0)|_{\td{X}} \oplus {\cal O}_{{\cal A}} (1, 0, 3)|_{\td{X}}\,. 
\label{A4}
\ee
We now want to restrict the sequence~\eqref{A2} to the curve $C_1$. 
For the curve of the form~\eqref{A1}, we obtain 
\be 
T {\cal A}|_{C_1} = {\cal O}_{C_1} (2) \oplus {\cal O}_{C_1}^4\,, \qquad 
N\td{X}|_{C_1}=  {\cal O}_{C_1} (1) \oplus {\cal O}_{C_1} (1)\,. 
\label{A5}
\ee
The sequence~\eqref{A2} then becomes
\be
0 \longrightarrow T\td{X}|_{C_1}  \stackrel{h_2|_{C_1} }{\longrightarrow}  {\cal O}_{C_1} (2) \oplus {\cal O}_{C_1}^4 
\stackrel{h_1|_{C_1}}{\longrightarrow}  {\cal O}_{C_1} (1) \oplus {\cal O}_{C_1} (1) \longrightarrow 0\,. 
\label{A6}
\ee
Let us now analyze the maps $h_1$ and $h_2$. The map $h_1$ is defined as a map from tangent directions $\pt$ along 
${\cal A}$ to the column vector $(\pt F_1, \pt F_2)^T$. Since $T{\cal A}$ is of rank 5  and $N\td{X}$ is of rank 2, $h_1$ is a 
$2 \times 5$ matrix. 
Evaluating the derivatives of $F_1$ and $F_2$ and restricting the results to the curve~\eqref{A1} gives
\begin{eqnarray}
h_1|_{C_1}  = \begin{pmatrix} 0 & 3 t_0 & t_1 & 0 & 0   \\ 0 & 0 & 0& 3(\l_1 t_0+ t_1) & \l_2 t_0 + \l_3 t_1   \end{pmatrix} \,.
\label{A7}
\end{eqnarray}
Since the sequence~\eqref{A6} is exact, it follows that $h_1$ and $h_2$ satisfy the composition rule $h_1 \circ h_2=0$. 
This determines $h_2|_{C_1} $ to be 
\begin{eqnarray}
h_2|_{C_1}  = \begin{pmatrix} 1 & 0 &  0   \\ 0 & t_1 & 0 \\ 0 & -3 t_0 & 0 \\ 0 & 0 &  \l_2 t_0 + \l_3 t_1 \\ 0 & 0 &  -3(\l_1 t_0+ t_1) \end{pmatrix} 
\label{A8}
\end{eqnarray}
up to an arbitrary holomorphic section $h_0$ on $C_1$ which is a homogeneous polynomials of degree $k\geq 0$ in $[t_0: t_1]$. 
Since any vector bundle on $C_1 \simeq {\mathbb P}^1$ is a sum of line bundles, $T\td{X}|_{C_1}$ must be of the form 
${\cal O}_{C_1} (m_1) \oplus {\cal O}_{C_1} (m_2) \oplus {\cal O}_{C_1} (m_3)$ where, from~\eqref{A6}, it follows that $m_1+m_2+m_3=0$. 
Examining the sequence~\eqref{A6}, it is easy to see that these conditions have  only one consistent possibility, namely $m_1=2, m_2=m_3=-1$. 
This means that
\be
T\td{X}|_{C_1} = {\cal O}_{C_1} (2) \oplus {\cal O}_{C_1} (-1) \oplus {\cal O}_{C_1} (-1)\,. 
\label{A9}
\ee
Finally, let us consider the short exact sequence relating the tangent bundle $TC_1$ and the normal bundle $N C_1$  of $C_1$ given by
\be
0 \longrightarrow TC_1 \longrightarrow T \td{X}|_{C_1} \longrightarrow N C_1 \longrightarrow 0\,. 
\label{A10}
\ee
Using eq.~\eqref{A9}, we obtain
\be
0 \longrightarrow {\cal O}_{C_1} (2)  \longrightarrow {\cal O}_{C_1} (2) \oplus {\cal O}_{C_1} (-1) \oplus {\cal O}_{C_1} (-1) \longrightarrow N C_1 \longrightarrow 0\,. 
\label{A11}
\ee
This implies that  the only possible form for $NC_1$ is
\be 
NC_1 =  {\cal O}_{C_1} (-1) \oplus {\cal O}_{C_1} (-1)\,. 
\label{A12}
\ee
%


\section{Extension of $\td{W}$ and $\td{V}$}
\label{B}


In this appendix, we calculate the number of extensions of $\td{W}$ and $\td{V}$ and
prove eq.~\eqref{3.8}. Our calculations will be similar to the ones performed in~\cite{Braun:2005zv}, where additional details can be found. 

\subsection{Extensions of  $\td{W}$}


The extensions of $\td{W}$ are given by the dimension of the cohomology group 
\be
H^1 (\td{X}, L_1 \otimes L_2^*)= H^1 (\td{X}, {\cal O}_{\td{X}} (-2 \phi + \t_1+ 2 \t_2))\,. 
\label{B1}
\ee
Let us consider the direct image  $\pi_{1*} L_1 \otimes L_2^*$ under the projection $\pi_1$ in the diagram~\eqref{2.8}. Using the definitions
of $\phi, \t_1, \t_2$ in~\eqref{2.15}, we can give $L_1 \otimes L_2^*$ in the form 
\be 
L_1 \otimes L_2^* = \pi_1^* {\cal O}_{B_1} (t-2 f) \otimes \pi_2^* {\cal O}_{B_2} (2t)\,. 
\label{B2}
\ee
From the diagram~\eqref{2.8}, it follows that the projections satisfy 
\be 
\pi_{1*} \pi_2^* = \b_1^* \b_{2 *}\,. 
\label{B3}
\ee
Then we obtain 
\be 
\pi_{1* } L_1 \otimes L_2^* = {\cal O}_{B_1} (t-2 f) \otimes \b_1^* \b_{2 *} {\cal O}_{B_2} (2t) = 
\b_1^*  {\cal O}_{{\mathbb P}^1} (-2) \otimes {\cal O}_{B_1} (t)  \otimes \b_1^* \b_{2 *} {\cal O}_{B_2} (2t)\,. 
\label{B4}
\ee
Computing $R^1 \pi_{1*} L_1 \otimes L_2^*$, we find that
\be
R^1 \pi_{1*} L_1 \otimes L_2^* =0\,. 
\label{B5}
\ee
To show this, note that at each point $p$ on $B_1$, $R^1 \pi_{1*} L_1 \otimes L_2^*$ is generated by the first cohomology group
\be 
H^1 (F_p, {\cal O}_{\td{X}} (-2 \phi + \t_1+ 2 \t_2)|_{F_p})
\label{B6}
\ee
on the elliptic fiber $F_p$ of the projection $\pi_1$ at $p$. 
Using eqs.~\eqref{2.13}, \eqref{2.15} we find that the line bundle  ${\cal O}_{\td{X}} (-2 \phi + \t_1+ 2 \t_2)|_{F_p}$ has degree 3 
and by the Kodaira vanishing theorem (see e.g~\cite{GH}) the cohomology group in~\eqref{B6} vanishes. This proves~\eqref{B5}. 
As the next step we similarly project~\eqref{B4} to the base of $B_1$. We obtain 
\bea
&& 
\b_{1 *}\pi_{1*} L_1 \otimes L_2^* = {\cal O}_{{\mathbb P}^1} (-2) \otimes \b_{1 *}  {\cal O}_{B_1} (t) \otimes \b_{2 *} {\cal O}_{B_2} (2t)\,, 
\label{B7} \\
&& 
R^1 \b_{1 *}\pi_{1*} L_1 \otimes L_2^*=0\,. 
\label{B8}
\eea
Using the identities~\cite{Braun:2005zv}
\be 
 \b_{k *}  {\cal O}_{B_k} (t) =  {\cal O}_{{\mathbb P}^1}^{\oplus 3}\,, \quad 
 \b_{k *}  {\cal O}_{B_k} (2t) =  {\cal O}_{{\mathbb P}^1}^{\oplus 6}\,, \qquad k=1, 2
\label{B9}
\ee
we find that 
\be 
\b_{1 *}\pi_{1*} L_1 \otimes L_2^* = {\cal O}_{{\mathbb P}^1}(-2)^{\oplus 18}\,. 
\label{B10}
\ee
Since the higher direct images in eqs.~\eqref{B5}, \eqref{B8} vanish from a Leray spectral sequence, it follows that 
\be
h^1 (\td{X}, L_1 \otimes L_2^*) = h^1 ({\mathbb P}^1, \b_{1 *}\pi_{1*} L_1 \otimes L_2^*) =h^1 ({\mathbb P}^1,  {\cal O}_{{\mathbb P}^1}(-2)^{\oplus 18}) =18\,. 
\label{B11}
\ee
%


\subsection{Extensions of  $\td{V}$}


The number of extensions of $\td{V}$ (for a fixed extension $\td{W}$ in $[\td{W}]$) is given by $H^1 (\td{X}, \td{W} \otimes L_3^*)$. To compute this cohomology group, 
we consider the short exact sequence
\be 
0 \longrightarrow L_1\otimes L_3^* \longrightarrow \td{W} \otimes L_3^* \longrightarrow L_2 \otimes L_3^* \longrightarrow 0\,, 
\label{B12}
\ee
where 
\be 
L_1\otimes L_3^* ={\cal O}_{\td{X}} (-4 \phi + 5\t_1+  \t_2)\,, \quad 
L_2\otimes L_3^* ={\cal O}_{\td{X}} (-2 \phi + 4\t_1 -  \t_2)\,. 
\label{B13}
\ee
The sequence~\eqref{B12} implies the following long exact sequence of cohomology groups
\bea
0  \longrightarrow & H^0 (\td{X}, L_1\otimes L_3^*) & \longrightarrow H^0( \td{X}, \td{W}\otimes L_3^*) 
\longrightarrow  H^0 (\td{X}, L_2\otimes L_3^*)  \longrightarrow
\nonumber \\
& H^1 (\td{X}, L_1\otimes L_3^*) & \longrightarrow H^1( \td{X}, \td{W}\otimes L_3^*) 
\longrightarrow  H^1 (\td{X}, L_2\otimes L_3^*)  \longrightarrow
\nonumber \\
& H^2 (\td{X}, L_1\otimes L_3^*) & \longrightarrow \dots \,. 
\label{B14}
\eea
The cohomology of $L_1\otimes L_3^*$ and $L_2\otimes L_3^*$ can be computed using direct images, just as in the previous subsection. 
Using the identities~\cite{Braun:2005zv}
\bea
&&
 \b_{k *}  {\cal O}_{B_k} (4 t) =  {\cal O}_{{\mathbb P}^1}^{\oplus 9} \oplus 
{\cal O}_{{\mathbb P}^1}(1)^{\oplus 3} \,,
\nonumber \\
&& 
 \b_{k *}  {\cal O}_{B_k} (5 t) =  {\cal O}_{{\mathbb P}^1}^{\oplus 9} \oplus 
{\cal O}_{{\mathbb P}^1}(1)^{\oplus 6} \,, 
\nonumber \\
&& 
R^1 \b_{k *}  {\cal O}_{B_k} (- t) = 
{\cal O}_{{\mathbb P}^1}(-1)^{\oplus 3} 
\label{B15}
\eea
and following the same steps as in the previous subsection, we obtain 
\bea
&& 
H^0 (\td{X}, L_1\otimes L_3^*) = H^2 (\td{X}, L_1\otimes L_3^*)=0\,, 
\nonumber \\
&&
H^1 (\td{X}, L_1\otimes L_3^*) = H^1 ({\mathbb P}^1, \b_{1 *} \pi_{1 *} L_1\otimes L_3^*)= 
H^1 ({\mathbb P}^1, 
{\cal O}_{{\mathbb P}^1}(-4)^{\oplus 27} \oplus 
{\cal O}_{{\mathbb P}^1}(-3)^{\oplus 18})\,, \nonumber \\
&&
 h^1 (\td{X}, L_1\otimes L_3^*) =117\,, 
\nonumber \\
&&
H^0 (\td{X}, L_2\otimes L_3^*) = H^1 (\td{X}, L_2\otimes L_3^*)= H^2 (\td{X}, L_2\otimes L_3^*)=0\,. 
\label{B18}
\eea
Then from~\eqref{B14} we see that 
\be
H^1( \td{X}, \td{W}\otimes L_3^*) = H^1 (\td{X}, L_1\otimes L_3^*) \,, \qquad h^1( \td{X}, \td{W}\otimes L_3^*) =117\,. 
\label{B19}
\ee
%


\section{Stability of $\td{W}$ and $\td{V}$}
\label{C}


Since we are only considering a toy model, we will not give a comprehensive proof that $\td{W}$ and $\td{V}$ are stable. 
Instead, we examine the most important necessary condition for  this to be the case.

Let us recall that a vector bundle  $\td{{V}}$ on $\td{X}$ is called stable if for any subsheaf ${S}$ of lower rank we have 
\be 
\mu ({S}) < \mu (\td{{V}}) \,. 
\label{C1}
\ee
Here, the slope $ \mu ({S}) $ is defined by 
\be
\mu ({S}) = \frac{1}{{\rm rk} ({S})} \int_{\td{X}} c_1 ({S}) \wedge \o_{\td{X}} \wedge \o_{\td{X}}\,, 
\label{C2}
\ee
where $\o_{\td{X}}$ is the Kahler form on $\td{X}$.

From eqs.~\eqref{3.3}, we observe that the line bundle $L_1$ injects into $\td{W}$ and $\td{W}$ injects into $\td{V}$. 
We now discuss whether   $L_1$ and  $\td{W}$ destabilize $\td{W}$ and $\td{V}$ respectively. 
Using the definition of $\td{W}$ in~\eqref{3.3}, we see that $\td{W}$  has rank 2 and its first Chern class
is given by $c_1 (L_1) + c_1 (L_2)$. Then the condition $\mu (L_1) < \mu (\td{W})$ can be stated as
\be 
\int_{\td{X}} (c_1 (L_1) -  c_1 (L_2))  \wedge \o_{\td{X}} \wedge \o_{\td{X}} <0 ~~~\Leftrightarrow~~~  \mu (L_1 \otimes L_2^*) <0\,. 
\label{C3}
\ee
Since $L_1$ and $L_2$ are equivariant and constructed out of the invariant classes, we can replace  $\o_{\td{X}}$ in~\eqref{C3}
with its invariant part $\o_X$ in~\eqref{2.17}.
Using the expression for the invariant part of the Kahler form in~\eqref{2.17}, we can rewrite~\eqref{C3} in the form
\be 
\int_{\td{X}} (-2 \o_{\phi} + \o_{\t_1} + 2 \o_{\t_2}) \wedge (t^1 \o_{\phi} + t^2 \o_{\t_1} + t^3 \o_{\t_2})^2 <0\,. 
\label{C4}
\ee
Using the triple intersection numbers~\eqref{2.16}, we  then obtain the following inequality for the Kahler parameters:
\be (t^3)^2 + 4 (t^1)^2 + 6 t^1 t^3+ 24 t^1 t^2 - 6 t^2 t^3 <0 \,.
\label{C5}
\ee

Let us now study the condition that $\mu (\td{W}) < \mu(\td{V})$. Note that $c_1 (\td{W})= c_1 (L_1) + c_1 (L_2)$
and, since $L_1 \otimes L_2 \otimes L_3$ is trivial, it follows that $c_1 (\td{W}) =c_1 (L_3^*)$. Also note that since 
$c_1 (\td{V})=0$, it  follows that $\mu (\td{V}) =0$. Then the condition $\mu (\td{W}) < \mu(\td{V})$ can be stated as
\be 
\mu (L_3^*) <0\, ~~~\Leftrightarrow~~~
\int_{\td{X}} (-2 \o_{\phi} + 3\o_{\t_1} ) \wedge (t^1 \o_{\phi} + t^2 \o_{\t_1} + t^3 \o_{\t_2})^2 <0\,. 
\label{C6}
\ee
Using the triple intersection numbers~\eqref{2.16}, we then obtain the inequality
\be
(t^3)^2 + 6 t^1 t^3 - 4 t^2 t^3 <0\,. 
\label{C7}
\ee
The bundles $\td{W}$ and $\td{V}$ are not destabilized if there exists a region in the Kahler moduli space 
where both inequalities~\eqref{C5} and~\eqref{C7} are simultaneously satisfied. It is easy to see that it is indeed the case.
For example, if we take  $t^2 \approx t^3$ and $t^1 \ll t^2, t^3$ both inequalities are satisfied. 


\section{Parameterization of  the moduli space of $V$}
\label{D}


Let us recall from section~\ref{bundle} that the invariant extensions in $[\td{V}]$, as well as the space of
extensions $[V]$, are described by the invariant subspace of the quotient 
\be
\dfrac{H^1 ({\cal A}, {\cal O}_{{\cal A}} (-4, 5, 1))}{F_1 \cdot H^1 ({\cal A}, {\cal O}_{{\cal A}} (-5, 2, 1))}\,. 
\label{D1}
\ee
The elements of the numerator were parameterized as
\be 
v_{inv} = r_0^2 f_1 ({\bf x}, {\bf y}) +r_0 r_1 f_2 ({\bf x}, {\bf y}) + r_1^2 f_3 ({\bf x}, {\bf y})\,, 
\label{D2}
\ee
where $f_1, f_2, f_3$ are invariant polynomials of degree $(5,1)$ on ${\mathbb P}^2 \times {\mathbb P}^2$
and $\{r_0, r_1\}$ is a basis in  the vector space $H^1 ({\mathbb P}^1, {\cal O}_{{\mathbb P}^1} (-3))$ dual to the 
basis $\{t_0, t_1\}$  in $H^0 ({\mathbb P}^1, {\cal O}_{{\mathbb P}^1} (1))$. The polynomials $f_1, f_2, f_3$
can be expanded in the basis~\eqref{3.29} 
\be 
f_1= \sum_{\a=1}^7 a_{\a} E_{\a}\,, \qquad 
f_2= \sum_{\a=1}^7 b_{\a} E_{\a}\,, \qquad f_3= \sum_{\a=1}^7 c_{\a} E_{\a}\,. 
\label{D3}
\ee
The aim of this appendix is to describe the process of factoring out  $F_1 \cdot H^1 ({\cal A}, {\cal O}_{{\cal A}} (-5, 2, 1))$. 
This will give a parameterization of the invariant part of the moduli space of $\td{V}$ and of the 
moduli space of $V$. 

Consider an element $u$ in $H^1 ({\cal A}, {\cal O}_{{\cal A}} (-5, 2, 1))$. Let us write it in the form similar to~\eqref{D2}. 
Using the Kunneth  and  Bott formulas, we can express $H^1 ({\cal A}, {\cal O}_{{\cal A}} (-5, 2, 1))$ as
\be
H^1 ({\cal A}, {\cal O}_{{\cal A}} (-5, 2, 1)) = H^1({\mathbb P}^1, {\cal O}_{{\mathbb P}^1} (-5) )\otimes
H^0({\mathbb P}^2 \times {\mathbb P}^2 , {\cal O}_{{\mathbb P}^2 \times {\mathbb P}^2} (2, 1)) \,. 
\label{D4}
\ee
In the first factor 
\be 
H^1({\mathbb P}^1, {\cal O}_{{\mathbb P}^1} (-5) ) \simeq H^0({\mathbb P}^1, {\cal O}_{{\mathbb P}^1} (3) )^*
\label{D5}
\ee
we can introduce a natural basis $\{r_0^3, r_0^2 r_1, r_0 r_1^2, r_1^3\}$ dual to the basis $\{t_0^3, t_0^2 t_1, t_0 t_1^2, t_1^3\}$
of homogeneous polynomials of degree 3 in $H^0({\mathbb P}^1, {\cal O}_{{\mathbb P}^1} (3) )$. 
Then $u$ can be written as
\be
u=  r_0^3 g_1 ({\bf x}, {\bf y})+ r_0^2 r_1 g_2({\bf x}, {\bf y}) + r_0 r_1^2 g_3 ({\bf x}, {\bf y})+ r_1^3 g_4({\bf x}, {\bf y})\,, 
\label{D6}
\ee
where $g_1, g_2, g_3, g_4$ are homogeneous polynomials of degree $(2, 1)$ on ${\mathbb P}^2 \times {\mathbb P}^2$. 
To restrict to invariant elements $u_{inv}$, we take $g_1, g_2, g_3, g_4$ to be invariant polynomials. 
The basis of invariant polynomials of degree $(2, 1)$ can be chosen to be 
\bea
&&
e_1= x_0^2 y_0 +x_1^2 y_1 +x_2^2 y_2\,, 
\nonumber \\
&& 
e_2= x_1 x_2  y_0 +x_2  x_0 y_1 +x_0 x_1 y_2\,. 
\label{D7}
\eea

Now let us consider the map $F_1$. Using eq.~\eqref{2.3}, we can write it in the form 
\be 
F_1= t_0 \s_0 ({\bf x}) + t_1 \s_1 ({\bf x}) ~~~~{\rm where}~~~~ \s_0 ({\bf x}) =x_0 x_1 x_2\,, \  \s_1 ({\bf x}) =x_0^3+ x_1^3+  x_2^3\,. 
\label{D8}
\ee
Let us  multiply $u_{inv}$ by $F_1$, using the fact that the bases $\{r_0, r_1\}$ and $\{t_0, t_1\}$ are dual to each other. We obtain 
\be 
F_1 u_{inv}= r_0^2 (\s_0 g_1 + \s_1 g_2) +   r_0 r_1 (\s_0 g_2 + \s_1 g_3)  +   r_1^2 (\s_0 g_3 + \s_1 g_4)\,. 
\label{D9}
\ee
Comparing this to eq.~\eqref{D2}, we see that we have to mod out by the equivalence relations 
\be 
f_1 \sim f_1+  \s_0 g_1 + \s_1 g_2\,, \qquad f_2 \sim f_2+  \s_0 g_2 + \s_1 g_3 \,,\qquad f_3 \sim f_3+  \s_0 g_3 + \s_1 g_4\,.
\label{D10}
\ee
Now our aim is to represent $F_1$ in matrix form. From~\eqref{D9}, we see that we can write it as 
\begin{eqnarray}
F_1  = \begin{pmatrix} \s_0 & \s_1 & 0 & 0   \\ 0 & \s_0 & \s_1 & 0 \\ 0 & 0 & \s_0 & \s_1   \end{pmatrix} \,.
\label{D11}
\end{eqnarray}
This matrix acts on the column vector $(g_1, g_2, g_3, g_4)^T$. Let us now write this matrix in the bases 
$E_{\a}$ in~\eqref{3.29} and $e_{\b}$ in~\eqref{D7}. In this basis, $F_1$ is a $7\cdot 3 \times 2 \cdot 4 = 21 \times 8$ matrix. 
To fully express this matrix, we have to study the action of $\s_0$ and $\s_1$ on the basis polynomials in~\eqref{D7} and present the result in terms of the basis polynomials
in~\eqref{3.29}. It is straightforward to show that 
\bea
&&
\s_0  e_1  \equiv (\s_0)^{\a}_{\ 1}  E_{\a} = E_1 +E_2 +E_3 \,, \qquad \s_0  e_2  \equiv  (\s_0)^{\a}_{\ 2}  E_{\a} = E_4 +E_5 +E_6 \,,
\nonumber \\
&& 
\s_1  e_1  \equiv  (\s_1)^{\a}_{\ 1} E_{\a} = E_4 \,, \qquad \s_0  e_2  \equiv (\s_0)^{\a}_{\ 2}  E_{\a} = E_7\,. 
\label{D12}
\eea
This leads to the following matrices for $ (\s_0)^{\a}_{\ \b}$ and $ (\s_1)^{\a}_{\ \b}$:
\begin{eqnarray}
(\s_0)^{\a}_{\ \b}  = \begin{pmatrix}  1   & 0 \\ 1 & 0 \\ 1 & 0 \\ 0 & 1 \\  0 & 1 \\0 & 1 \\ 0 & 0 \end{pmatrix} \,, \qquad 
(\s_1)^{\a}_{\ \b}  = \begin{pmatrix}  0   & 0 \\ 0 & 0 \\ 0 & 0 \\ 1 & 0 \\  0 & 0 \\0 & 0 \\ 0 & 1 \end{pmatrix} \,.
\label{D13}
\end{eqnarray}
Inserting~\eqref{D13} into~\eqref{D11}, gives the full matrix $F_1$. 

Performing the quotient action in~\eqref{D1} is now equivalent to finding the cokernel of the matrix $F_1$ in~\eqref{D11}, \eqref{D13}. 
This is, in turn, equivalent to finding the kernel of the matrix $(F_1)^T$ which acts on the parameters of the polynomials
$f_1, f_2, f_3$; that is, on the column vector $(a_1, \dots, a_7, b_1, \dots, b_7, c_1, \dots, c_7)^T$.
Finding the kernel of $(F_1)^T$ means solving the linear system of equations
\be 
(F_1)^T (a_1, \dots, a_7, b_1, \dots, b_7, c_1, \dots, c_7)^T=( (a_1, \dots, a_7, b_1, \dots, b_7, c_1, \dots, c_7) F_1)^T=0 \ ,
\label{D14}
\ee
which is equivalent to 
\be 
(a_1, \dots, a_7, b_1, \dots, b_7, c_1, \dots, c_7) F_1 =0\,. 
\label{D15}
\ee
Using the matrix form of $F_1$ in eqs.~\eqref{D11} and \eqref{D13}, it is easy to see that the  system of linear equations~\eqref{D15}
in components becomes~\eqref{3.31}. 



\newpage
\begin{thebibliography}{99}

 
 \bibitem{Braun:2005ux}
  V.~Braun, Y.~H.~He, B.~A.~Ovrut and T.~Pantev,
  ``A Heterotic standard model,''
  Phys.\ Lett.\ B {\bf 618} (2005) 252
  [hep-th/0501070].
  
  
\bibitem{Braun:2005bw} 
  V.~Braun, Y.~H.~He, B.~A.~Ovrut and T.~Pantev,
  ``A Standard model from the E(8) x E(8) heterotic superstring,''
  JHEP {\bf 0506}, 039 (2005)
  [hep-th/0502155].
  
\bibitem{Braun:2005zv} 
  V.~Braun, Y.~H.~He, B.~A.~Ovrut and T.~Pantev,
  ``Vector bundle extensions, sheaf cohomology, and the heterotic standard model,''
  Adv.\ Theor.\ Math.\ Phys.\  {\bf 10}, no. 4, 525 (2006)
  [hep-th/0505041].
  
 \bibitem{Braun:2005nv} 
  V.~Braun, Y.~H.~He, B.~A.~Ovrut and T.~Pantev,
  ``The Exact MSSM spectrum from string theory,''
  JHEP {\bf 0605}, 043 (2006)
  [hep-th/0512177].
  
    \bibitem{Anderson:2007nc} 
  L.~B.~Anderson, Y.~H.~He and A.~Lukas,
  ``Heterotic Compactification, An Algorithmic Approach,''
  JHEP {\bf 0707}, 049 (2007)
  [hep-th/0702210 [HEP-TH]].
  
  
 \bibitem{Anderson:2009mh} 
  L.~B.~Anderson, J.~Gray, Y.~H.~He and A.~Lukas,
  ``Exploring Positive Monad Bundles And A New Heterotic Standard Model,''
  JHEP {\bf 1002}, 054 (2010)
  [arXiv:0911.1569 [hep-th]].
  
  \bibitem{Buchbinder:2013dna}
  E.~I.~Buchbinder, A.~Constantin and A.~Lukas,
  ``The Moduli Space of Heterotic Line Bundle Models: a Case Study for the Tetra-Quadric,''
  JHEP {\bf 1403} (2014) 025
  [arXiv:1311.1941 [hep-th]].
 
 \bibitem{Buchbinder:2014qda}
  E.~I.~Buchbinder, A.~Constantin and A.~Lukas,
  ``A heterotic standard model with $B - L$ symmetry and a stable proton,''
  JHEP {\bf 1406} (2014) 100
  [arXiv:1404.2767 [hep-th]].
 
 \bibitem{Buchbinder:2014sya}
  E.~I.~Buchbinder, A.~Constantin and A.~Lukas,
  ``Non-generic Couplings in Supersymmetric Standard Models,''
  Phys.\ Lett.\ B {\bf 748} (2015) 251
  [arXiv:1409.2412 [hep-th]].
 
\bibitem{Anderson:2010mh} 
  L.~B.~Anderson, J.~Gray, A.~Lukas and B.~Ovrut,
  ``Stabilizing the Complex Structure in Heterotic Calabi-Yau Vacua,''
  JHEP {\bf 1102}, 088 (2011)
  [arXiv:1010.0255 [hep-th]].
  
\bibitem{Anderson:2011cza} 
  L.~B.~Anderson, J.~Gray, A.~Lukas and B.~Ovrut,
  ``Stabilizing All Geometric Moduli in Heterotic Calabi-Yau Vacua,''
  Phys.\ Rev.\ D {\bf 83}, 106011 (2011)
  [arXiv:1102.0011 [hep-th]].
  
\bibitem{Dine:1986zy} 
  M.~Dine, N.~Seiberg, X.~G.~Wen and E.~Witten,
  ``Nonperturbative Effects on the String World Sheet,''
  Nucl.\ Phys.\ B {\bf 278}, 769 (1986).

\bibitem{Dine:1987bq} 
  M.~Dine, N.~Seiberg, X.~G.~Wen and E.~Witten,
  ``Nonperturbative Effects on the String World Sheet. 2.,''
  Nucl.\ Phys.\ B {\bf 289}, 319 (1987).
  
\bibitem{Becker:1995kb} 
  K.~Becker, M.~Becker and A.~Strominger,
  ``Five-branes, membranes and nonperturbative string theory,''
  Nucl.\ Phys.\ B {\bf 456}, 130 (1995)
  [hep-th/9507158].

\bibitem{Witten:1996bn} 
  E.~Witten,
  ``Nonperturbative superpotentials in string theory,''
  Nucl.\ Phys.\ B {\bf 474}, 343 (1996)
  [hep-th/9604030].
  
\bibitem{Donagi:1996yf} 
  R.~Donagi, A.~Grassi and E.~Witten,
  ``A Nonperturbative superpotential with E(8) symmetry,''
  Mod.\ Phys.\ Lett.\ A {\bf 11}, 2199 (1996)
  [hep-th/9607091].

\bibitem{Witten:1999eg} 
  E.~Witten,
  ``World sheet corrections via D instantons,''
  JHEP {\bf 0002}, 030 (2000)
  [hep-th/9907041].
  
\bibitem{Harvey:1999as} 
  J.~A.~Harvey and G.~W.~Moore,
  ``Superpotentials and membrane instantons,''
  hep-th/9907026.
  
\bibitem{Lima:2001jc} 
  E.~Lima, B.~A.~Ovrut, J.~Park and R.~Reinbacher,
  ``Nonperturbative superpotential from membrane instantons in heterotic M theory,''
  Nucl.\ Phys.\ B {\bf 614}, 117 (2001)
  [hep-th/0101049].

\bibitem{Lima:2001nh} 
  E.~Lima, B.~A.~Ovrut and J.~Park,
  ``Five-brane superpotentials in heterotic M theory,''
  Nucl.\ Phys.\ B {\bf 626}, 113 (2002)
  [hep-th/0102046].
  
\bibitem{Buchbinder:2002ic} 
  E.~I.~Buchbinder, R.~Donagi and B.~A.~Ovrut,
  ``Superpotentials for vector bundle moduli,''
  Nucl.\ Phys.\ B {\bf 653}, 400 (2003)
  [hep-th/0205190].
  
\bibitem{Buchbinder:2002pr} 
  E.~I.~Buchbinder, R.~Donagi and B.~A.~Ovrut,
  ``Vector bundle moduli superpotentials in heterotic superstrings and M theory,''
  JHEP {\bf 0207}, 066 (2002)
  [hep-th/0206203].
  
\bibitem{Beasley:2003fx} 
  C.~Beasley and E.~Witten,
  ``Residues and world sheet instantons,''
  JHEP {\bf 0310}, 065 (2003)
  [hep-th/0304115].

\bibitem{Beasley:2005iu} 
  C.~Beasley and E.~Witten,
  ``New instanton effects in string theory,''
  JHEP {\bf 0602}, 060 (2006)
  [hep-th/0512039].
  
 \bibitem{Schoen}
C.~Schoen, ``On fiber products of rational elliptic surfaces with section," {\it Math. Z.} 197 (1988), {\rm no.} 2, 177-199. 

\bibitem{Braun:2007tp} 
  V.~Braun, M.~Kreuzer, B.~A.~Ovrut and E.~Scheidegger,
  ``Worldsheet instantons, torsion curves, and non-perturbative superpotentials,''
  Phys.\ Lett.\ B {\bf 649}, 334 (2007)
  [hep-th/0703134].

\bibitem{Braun:2007xh} 
  V.~Braun, M.~Kreuzer, B.~A.~Ovrut and E.~Scheidegger,
  ``Worldsheet instantons and torsion curves, part A: Direct computation,''
  JHEP {\bf 0710}, 022 (2007)
  [hep-th/0703182 [HEP-TH]].
  
\bibitem{Braun:2007vy} 
  V.~Braun, M.~Kreuzer, B.~A.~Ovrut and E.~Scheidegger,
  ``Worldsheet Instantons and Torsion Curves, Part B: Mirror Symmetry,''
  JHEP {\bf 0710}, 023 (2007)
  [arXiv:0704.0449 [hep-th]].
  
  
\bibitem{Distler:1986wm} 
  J.~Distler,
  ``Resurrecting (2,0) Compactifications,''
  Phys.\ Lett.\ B {\bf 188}, 431 (1987).
  
\bibitem{Distler:1987ee} 
  J.~Distler and B.~R.~Greene,
  ``Aspects of (2,0) String Compactifications,''
  Nucl.\ Phys.\ B {\bf 304}, 1 (1988).
  
\bibitem{Silverstein:1995re} 
  E.~Silverstein and E.~Witten,
  ``Criteria for conformal invariance of (0,2) models,''
  Nucl.\ Phys.\ B {\bf 444}, 161 (1995)
  [hep-th/9503212].
  
\bibitem{Basu:2003bq} 
  A.~Basu and S.~Sethi,
  ``World sheet stability of (0,2) linear sigma models,''
  Phys.\ Rev.\ D {\bf 68}, 025003 (2003)
  [hep-th/0303066].
  
  \bibitem{Anderson:2008uw} 
  L.~B.~Anderson, Y.~H.~He and A.~Lukas,
  ``Monad Bundles in Heterotic String Compactifications,''
  JHEP {\bf 0807}, 104 (2008)
  [arXiv:0805.2875 [hep-th]].

\bibitem{Bott}
R.~Bott, L.~W.~Tu,  ``Differential Forms in Algebraic Topology," Springer 1982.

\bibitem{Aspinwall:1994uj} 
  P.~S.~Aspinwall and D.~R.~Morrison,
  ``Chiral rings do not suffice: N=(2,2) theories with nonzero fundamental group,''
  Phys.\ Lett.\ B {\bf 334}, 79 (1994)
  [hep-th/9406032].
  
\bibitem{Ovrut:2002jk} 
  B.~A.~Ovrut, T.~Pantev and R.~Reinbacher,
  ``Torus fibered Calabi-Yau threefolds with nontrivial fundamental group,''
  JHEP {\bf 0305}, 040 (2003)
  [hep-th/0212221].
  
 \bibitem{Persson}
U.~Persson, ``Configurations of Kodaira fibers on rational elliptic surfaces,"  
{\it Math. Z.} 205 (1990), {\rm no.} 1, 1-47.

   \bibitem{Shioda}
T.~Shioda, ``On the Mordell-Weil lattices," {\it Comment. Math. Univ. St.Paul.} {\bf 39} (1990), 
{\rm no.} 2, 211-240.

\bibitem{OS}
K.~Oguiso and T.~Shioda, ``The Mordell-Weil lattice of a rational elliptic surface," {\it Comment. Math. Univ. St.Paul.} {\bf 40} (1991), 
{\rm no.} 1, 83-99. 

\bibitem{Hubsch}
T.~Hubsch, ``Calabi-Yau Manifolds: A Bestiary For Physicists," {\it World Scientific Publishing} 1992.

\bibitem{Candelas:1987se}
P.~Candelas,  ``Yukawa Couplings Between (2,1) Forms,''
Nucl.\ Phys.\ B {\bf 298} (1988) 458.
  
\bibitem{Blesneag:2015pvz}
  S.~Blesneag, E.~I.~Buchbinder, P.~Candelas and A.~Lukas,
  ``Holomorphic Yukawa Couplings in Heterotic String Theory,''
  JHEP {\bf 1601} (2016) 152,
  [arXiv:1512.05322 [hep-th]].
  
\bibitem{Blesneag:2016yag} 
  S.~Blesneag, E.~I.~Buchbinder and A.~Lukas,
  ``Holomorphic Yukawa Couplings for Complete Intersection Calabi-Yau Manifolds,''
  arXiv:1607.03461 [hep-th].  
  
\bibitem{GH}
 P.~Griffiths, J.~Harris,
``Principles of Algebraic Geometry,"
 Wiley Classics Library, 2011.

  

  

  

  

  





  
  
 
 
 
 
 
 
 
 
 
 
 
 
 
 
  
  
  
  
  

\end{thebibliography}
\end{document}